# An Atom Counting QSPR Protocol

S. Giri[1], D. R. Roy[1], P. Bultinck[2,*], V. Subramanian[3,*] and P. K. Chattaraj[1,*]

[1]Department of Chemistry, Indian Institute of Technology, Kharagpur 721302, India

[2]Department of Inorganic and Physical Chemistry, Ghent University, Krijgslaan 281,B-9000 Gent, Belgium

[3]Chemical Laboratory, Central Leather Research Institute, Adyar, Chennai 600 020, India

## Abstract

A deceptively simple descriptor, viz. the number of carbon / non-hydrogenic atoms present in a molecule, is proposed for the development of useful quantitative-structure-property-relationship (QSPR) models. It is tested in models pertaining to the estimation of boiling point of alcohols, enthalpy of vaporization of polychlorinated biphenyls (PCBs), n-octanol / water partition coefficient of PCBs and chloroanisoles, p$K_a$ values of carboxylic acids, phenols and alcohols etc. Very high values of various regression coefficients ($R^2$, $R^2_{CV}$, $R^2_{Ad}$) suggest the significance of this descriptor which further improves in the resulting two-parameter QSPR models with electrophilicity or its local variant as an additional descriptor.

[*] To receive all correspondence: P. K. Chattaraj
    Department of Chemistry,
    Indian Institute of Technology,
    Kharagpur 721302, India
    E-mail: pkc@chem.iitkgp.ernet.in
    Fax: +91-3222-255303

**Introduction**

Quantitative Structure- Activity Relationship (QSAR) and Quantitative Structure- Property Relationship (QSPR) studies have seen recent upsurge of interest because of their potential in predicting various activities and properties of complex molecules bypassing the actual experimental observation. These studies rest on the presupposition that the molecular descriptors bear the signature of the molecular structure and hence are capable of providing insights into the molecular activity/reactivity and properties. However, it becomes a daunting task to find out the appropriate molecular descriptors for a given physical/chemical/biological property. Our aim is to provide with a simple descriptor, which does not need any large scale computation. We propose that the number of atoms in a molecule should serve that purpose. The rationale behind our prognosis is as follows. For a given group of molecules of similar structure the number of atoms will roughly scale as the number of electrons (N). It will also provide a rough estimate of the molecular volume (geometry) and hence the external potential ($v(\vec{r})$). Since N and $v(\vec{r})$ fix the Hamiltonian of the system they will provide the wave function and accordingly all properties.

In order to verify our prognosis we select some recently studied systems. We provide a QSPR protocol to use the number of atoms as the descriptor of boiling points of alcohol, enthalpy of vaporization of polychlorinated biphenyls (PCBs), n- octanol / water partition coefficients of PCBs and chloroanisoles and p$K_a$ values for carboxylic acids, phenols and alcohols. Since electrophilicity [1, 2] has been shown to be a reliable descriptor for these properties we also analyze the associated two- parameter (number of atoms and electrophilicity) QSPR models.

## QSPR Models for Boiling Points of Alcohols

Various QSPR models based on connectivity indices have been proposed [3-5] in recent times in estimating the boiling points of 58 aliphatic alcohols. We try to develop a model based on just the number of carbon atoms ($N_C$) present in each alcohol as the descriptor. It may be noted that no computation other than a simple atom counting is needed to obtain our model.

Table 1 compares the experimental boiling points of 58 aliphatic alcohols with the corresponding quantities calculated using $N_C$ as the sole descriptor. Figure 1 shows the associated regression with $R^2 = 0.864$, internal predictive ability ($R^2_{CV}$) = 0.853 and $R^2_{Ad}$ = 0.862. Considering the simplicity of the model it is unbelievable.

In order to improve upon this model we include electrophilicity ($\omega$) [1,2] as well so that we have a two-parameter model. Conceptual density functional theory [6-8] based reactivity descriptors like electronegativity ($\chi$) [9, 10], hardness ($\eta$) [11,12] and electrophilicity ($\omega = \frac{\chi^2}{2\eta}$) [1, 2] have been shown to be very useful in illucidating various aspects of structure, properties, reactivity and interactions [2, 6-8]. Here we use the following formula for the calculation of ($\omega$):

$$\omega = \frac{\chi^2}{2\eta}$$

$$\approx \frac{1}{4}\frac{(I+A)^2}{(I-A)} \approx \frac{1}{4}\frac{(\epsilon_{HOMO}+\epsilon_{LUMO})^2}{(\epsilon_{LUMO}-\epsilon_{HOMO})} \qquad (1)$$

where I, A, $\epsilon_{HOMO}$ and $\epsilon_{LUMO}$ are ionization potential, electron affinity, energy of the highest occupied molecular orbital and energy of the lowest unoccupied molecular orbital respectively. Necessary quantities are calculated within a B3LYP/6-31+G (d) level of theory.

Table 1 presents the calculated boiling points for this two parameter QSPR model using $N_C$, and $\omega$ as the descriptors, whereas Figure 2 highlights the beautiful correlation with $R^2 = 0.869$, $R^2_{CV} = 0.858$ and $R^2_{Ad} = 0.867$.

## QSPR Models for Enthalpy of Vaporization of PCBs

Since the number of carbon atoms in 209 PCB congeners (See the template in Table 2) remains same we use the number of non-hydrogenic atoms ($N_{NH}$) as the descriptor for the enthalpy of vaporization of these PCB molecules [13-16]. Experimental $\Delta_{vap}H_m$ values are available in two references [14, 15] and two different models are proposed based on those values as was done in reference 13. Tables 2 and 3 present the experimental and calculated $\Delta_{vap}H_m$ values for 17 and 27 congeners with $R^2$, $R^2_{(CV)}$, $R^2_{(Ad)}$ values as 0.908, 0.866, 0.902 and 0.853, 0.832, 0.847 respectively. Related plots are provided in Figures 3 and 4. Corresponding two parameter models are developed using $N_{NH}$ and electrophilicity ($\omega$) values. Necessary $\omega$ values are taken from reference 13. As is evidenced from Tables 2, 3 and Figures 3, 4 there exist beautiful linear correlations with $R^2$, $R^2_{(CV)}$, $R^2_{(Ad)}$ values as 0.951, 0.935, 0.948 and 0.856, 0.836, 0.851 respectively.

Tables 4, 5 and Figure 5 provide the predicted $\Delta_{vap}H_m$ values through the QSPR models presented above, for the remaining PCBs for which the corresponding experimental values are not known. These values compare very well with other reported literature values [13, 16].

# QSPR Models for n- Octanol/Water Partition Coefficients of PCBs and Chloroanisoles

There have been several attempts to develop QSPR models for the lipophilic behavior of PCBs [17] and chloroanosoles [18]. In the present work we provide one and two-parameter models with $N_{NH}$ and ($N_{NH}$, $\omega$) as descriptors. Required $\omega$ values are taken from references 17 and 18.

## Lipophilicity of PCBs

From a data set of 133 PCB congeners a training set containing 100 molecules is formed. Both single and two-parameter models are used to calculate the log P values of 100 PCBs, which are compared with the available experimental values taken from references 19 and 20. Figure 6 depicts the beautiful correlation with $R^2$, $R^2_{(CV)}$, $R^2_{(Ad)}$ as 0.832, 0.824, 0.830 and 0.905, 0.900, 0.905 respectively for $N_{NH}$ and ($N_{NH}$, $\omega$) as descriptors. These regression models are then used to predict the log P values of 33 PCBs constituting the test set. Table 7 and Figure 7 highlight the reliability and robustness of these models through the respective regression coefficients $R^2$, $R^2_{(CV)}$, $R^2_{(Ad)}$ being 0.873, 0.856, 0.869 and 0.906, 0.891, 0.903 respectively for one ($N_{NH}$)- and two ($N_{NH}$, $\omega$)- parameter models. This fact indirectly provides a justification for log P itself being a suitable QSAR/QSPR descriptor. For the whole data set of 133 PCB congeners these values are (Table 8, Figure 8), 0.842. 0.836, 0.841 and 0.906, 0.906, 0.906 respectively for one ($N_{NH}$)- and two ($N_{NH}$, $\omega$)- parameter models.

## Lipophilicity of Chloroanisoles

Several QSPR studies [18] have been carried out in order to estimate the log P values for chloroanisoles. In this work we try to achieve the same with one- and two-parameter QSPR models respectively with $N_{NH}$ and ($N_{NH}$, $\omega$) as the descriptors.

Table 9 compares the experimental log P values (Reference 21) of chloroanisoles with the corresponding values calculated using $N_{NH}$ as well as ($N_{NH}$, $\omega$) as descriptors. The $R^2$, $R^2_{(CV)}$, $R^2_{(Ad)}$ values (Table 9, Figure 9) are 0.927, 0.914, 0.923 and 0.928, 0.914, 0.924 respectively for one and two-parameter models. Unlike the case of PCBs there is very little improvement in going from a single parameter to a two-parameter case. The reason is the presence of a strong inter-correlation between $N_{NH}$ and $\omega$ with $R^2$, $R^2_{(CV)}$, $R^2_{(Ad)}$ as 0.957, 0.949, 0.955 (not shown here).

## QSPR Models for p$K_a$ Prediction

Group Philicity [22] has been found to be a reliable descriptor for p$K_a$ prediction. It is calculated as the condensed philicity [23] summed over a group of relevant atoms as

$$\omega_g^\alpha = \sum_{k=1}^{n} \omega \cdot f_k^\alpha \qquad (2)$$

where $f_k^\alpha$ is the condensed Fukui function on the atom k and $\alpha$ = +, -, 0 represents nucleophilic, electrophilic and radical attacks respectively. For the present work the required $\omega_g^+$ values are taken from reference 22 while the experimental p$K_a$ values are taken from references 24-26. Table 10 presents the experimental p$K_a$ values for three different sets of molecules comprising carboxylic acid, substituted phenols and alcohols along with their p$K_a$ values estimated in terms of $N_C$ and ($N_C$, $\omega_g^+$) via separate regressions for three sets. Figure 10 provides the linear correlation between the experimental and calculated p$K_a$ values. The correlation is very good with $R^2$, $R^2_{(CV)}$, $R^2_{(Ad)}$ being 0.9604, 0.9558, 0.9596 and 0.9906, 0.9895, 0.9904 respectively for the one ($N_C$)- and two ($N_C$, $\omega_g^+$)- parameter models. It is worth mentioning that in both the plots the slope is unity and the intercept is close to zero, as expected. Further work related to toxicity and and biological activity is in progress in our laboratory.

**Conclusions**

The number of atoms in a molecule can be considered to be a valid descriptor in a QSPR model. Very high coefficient of determination ($R^2$), leave-one-out cross-validated squared correlation coefficient ($R^2_{CV}$) and squared adjusted correlation coefficient ($R^2_{Ad}$) values associated with the QSPR models for estimating a variety of properties encompassing boiling points of alcohols, enthalpy of vaporization of PCBs, lipophilicity of PCBs and chloroanisoles and p$K_a$ values for carboxylic acids, phenols, and alcohols etc. highlight the reliability and robustness of these models and importance of a simple descriptor like the number of carbon/non-hydrogenic atoms in a molecule. Inclusion of electrophilicity or its local variant provides further improvement and a very versatile two-parameter QSPR model results.

**Acknowledgements**
We thank BRNS, Mumbai for financial assistance and Mr. J. Padmanabhan and Mr. R. Parthasarathi for their help in various ways.

**Table 1**. QSPR models for boiling points of 58 aliphatic alcohols.

| Alcohols | $\omega$ (eV) | Boiling Point | | |
|---|---|---|---|---|
| | | Exptl[*] | Calcd[a] ($N_C$) | Calcd[b] ($\omega$, $N_C$) |
| Methanol | 1.911 | 64.7 | 49.422 | 46.789 |
| Ethanol | 1.806 | 78.3 | 67.087 | 68.543 |
| 1-Propanol | 1.862 | 97.2 | 84.753 | 83.916 |
| 1-Butanol | 1.789 | 117 | 102.418 | 104.434 |
| 1-Pentanol | 1.896 | 137.8 | 120.084 | 117.761 |
| 1-Hexanol | 1.790 | 157 | 137.749 | 139.591 |
| 1-Heptanol | 1.790 | 176.3 | 155.415 | 157.193 |
| 1-Octanol | 1.788 | 195.2 | 173.080 | 174.856 |
| 1-Nonanol | 1.902 | 213.1 | 190.746 | 187.951 |
| 1-Decanol | 1.903 | 230.2 | 208.411 | 205.498 |
| 2-Propanol | 1.747 | 82.3 | 84.753 | 88.498 |
| 2-Butanol | 1.720 | 99.6 | 102.418 | 107.169 |
| 2-Pentanol | 1.692 | 119 | 120.084 | 125.886 |
| 2-Hexanol | 1.691 | 139.9 | 137.749 | 143.533 |
| 2-Octanol | 1.687 | 179.8 | 173.080 | 178.866 |
| 2-Nonanol | 1.686 | 198.5 | 190.746 | 196.508 |
| 3-Pentanol | 1.833 | 115.3 | 120.084 | 120.269 |
| 3-Hexanol | 1.798 | 135.4 | 137.749 | 139.268 |
| 3-Heptanol | 1.775 | 156.8 | 155.415 | 157.779 |
| 4-Heptanol | 1.793 | 155 | 155.415 | 157.045 |
| 3-Nonanol | 1.787 | 194.7 | 190.746 | 192.520 |
| 4-Nonanol | 1.796 | 193 | 190.746 | 192.158 |
| 5-Nonanol | 1.767 | 195.1 | 190.746 | 193.296 |
| 2-M-1-Propanol | 1.863 | 107.9 | 102.418 | 101.502 |
| 2-M-2-Propanol | 1.895 | 82.4 | 102.418 | 100.225 |
| 2-M-1-Butanol | 1.867 | 128.7 | 120.084 | 118.924 |
| 2-M-2-Butanol | 1.886 | 102 | 120.084 | 118.188 |
| 3-M-1-Butanol | 1.823 | 131.2 | 120.084 | 120.660 |
| 3-M-2-Butanol | 1.780 | 111.5 | 120.084 | 122.386 |
| 2-M-1-Pentanol | 1.921 | 148 | 137.749 | 134.365 |
| 3-M-1-Pentanol | 1.918 | 152.4 | 137.749 | 134.485 |
| 4-M-1-Pentanol | 1.903 | 151.8 | 137.749 | 135.075 |
| 2-M-2-Pentanol | 1.863 | 121.4 | 137.749 | 136.701 |
| 3-M-2-Pentanol | 1.777 | 134.2 | 137.749 | 140.093 |
| 4-M-2-Pentanol | 1.802 | 131.7 | 137.749 | 139.120 |
| 2-M-3-Pentanol | 1.823 | 126.6 | 137.749 | 138.281 |
| 3-M-3-Pentanol | 1.882 | 122.4 | 137.749 | 135.909 |
| 2-M-2-Hexanol | 1.864 | 142.5 | 155.415 | 154.242 |
| 3-M-3-Hexanol | 1.850 | 142.4 | 155.415 | 154.782 |
| 7-M-1-Octanol | 1.793 | 206 | 190.746 | 192.255 |
| 2-E-1-Butanol | 1.811 | 146.5 | 137.749 | 138.749 |
| 3-E-3-Pentanol | 1.879 | 142.5 | 155.415 | 153.629 |
| 2-E-1-Hexanol | 1.834 | 184.6 | 173.080 | 173.016 |
| 2,2-M-M-1-Propanol | 1.925 | 113.1 | 120.083 | 116.609 |
| 2,2-M-M-1-Butanol | 1.883 | 136.8 | 137.749 | 135.879 |
| 2,3-M-M-1-Butanol | 1.874 | 149 | 137.749 | 136.229 |
| 3,3-M-M-1-Butanol | 1.882 | 143 | 137.749 | 135.930 |
| 2,3-M-M-2-Butanol | 1.882 | 118.6 | 137.749 | 135.924 |
| 3,3-M-M-2-Butanol | 1.858 | 120 | 137.749 | 136.874 |
| 2,3-M-M-2-Pentanol | 1.869 | 139.7 | 155.415 | 154.028 |
| 3,3-M-M-2-Pentanol | 1.849 | 133 | 155.415 | 154.849 |
| 2,2-M-M-3-Pentanol | 1.907 | 136 | 155.415 | 152.530 |
| 2,4-M-M-3-Pentanol | 1.893 | 138.8 | 155.415 | 153.085 |
| 2,6-M-M-4-Heptanol | 1.914 | 178 | 190.746 | 187.473 |
| 2,3-M-M-3-Pentanol | 1.911 | 139 | 155.415 | 152.378 |
| 3,5-M-M-4-Heptanol | 1.837 | 187 | 190.746 | 190.499 |
| 2,2,3-M-M-M-3-Pentanol | 1.886 | 152.2 | 173.080 | 170.979 |
| 3,5,5-M-M-M-1-Hexanol | 1.852 | 193 | 190.746 | 189.921 |

*Experimental data taken from References [3, 4, 5].
[a]Calculated boiling point values using one parameter ($N_C$) regression.
[b]Calculated boiling point values using two parameter ($\omega$, $N_C$) regression.

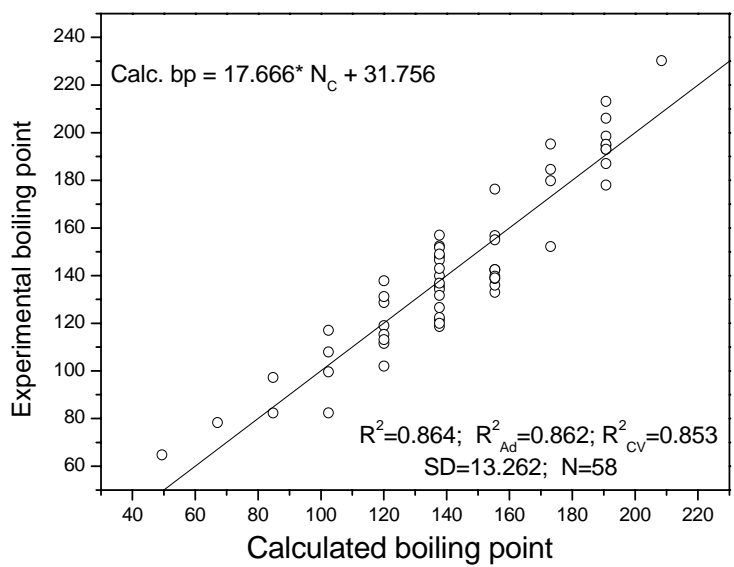

**Figure 1:** Experimental vs calculated boiling points with $N_C$ for the complete set of 58 aliphatic alcohols.

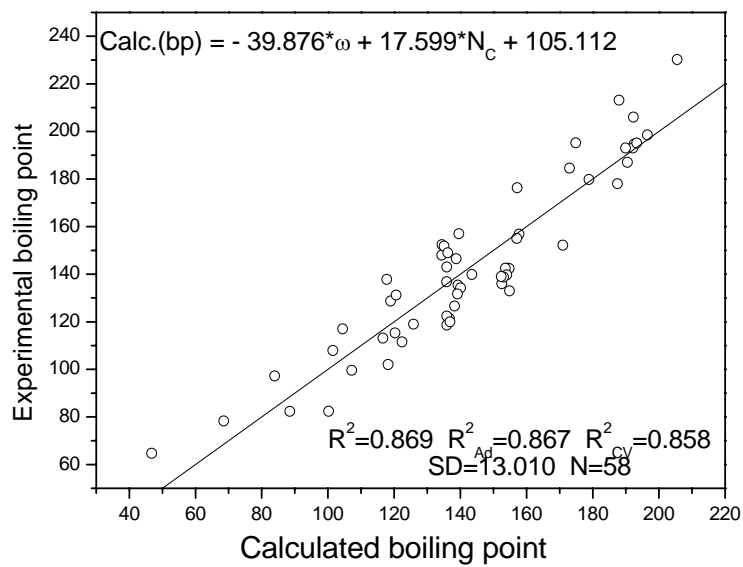

**Figure 2:** Experimental vs calculated boiling points with $N_C$, $\omega$ for the complete set of 58 aliphatic alcohols

**Regression Equations:**

| Parameters | Regression Equations | $R^2$ | $R^2_{(CV)}$ | $R^2_{(Ad)}$ |
|---|---|---|---|---|
| One | Calc. (bp) = 17.666* $N_C$ + 31.756 | 0.864 | 0.853 | 0.862 |
| Two | Calc. (bp) = - 39.876*$\omega$ + 17.599*$N_C$ + 105.112 | 0.869 | 0.858 | 0.867 |

**Table 2**. Enthalpy of vaporization for the data set of 17 Polychlorinated Biphenyls

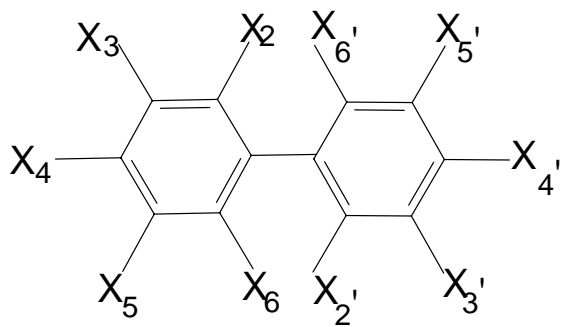

X, X′=H, or Cl
Geometry of biphenyl template with atom numbering

| IUPAC No | Compounds | $\omega$ | ($\Delta_{vap}H_m$ (298.15 K) | | |
|---|---|---|---|---|---|
| | | | Exptl* | Calcd[a] ($N_{NH}$) | Calcd[b] ($\omega$, $N_{NH}$) |
| | Biphenyl | 0.077 | 64.5± 2.2 | 65.21 | 62.45 |
| 1 | 2-Chlorobiphenyl | 0.087 | 72.1± 2.0 | 71.28 | 71.88 |
| 2 | 3-Chlorobiphenyl | 0.085 | 74.3± 1.1 | 71.28 | 70.57 |
| 3 | 4-Chlorobiphenyl | 0.085 | 71.6± 0.7 | 71.28 | 70.57 |
| 7 | 2,4-Dichlorobiphenyl | 0.096 | 75.4± 1.5 | 77.34 | 80.66 |
| 9 | 2,5-Dichlorobiphenyl | 0.095 | 76.8± 0.4 | 77.34 | 80.00 |
| 11 | 3,3'-Dichlorobiphenyl | 0.099 | 81± 0.2 | 77.34 | 82.62 |
| 15 | 4,4'-Dichlorobiphenyl | 0.090 | 81.4± 0.3 | 77.34 | 76.73 |
| 18 | 2,2',5-Trichlorobiphenyl | 0.094 | 80.2± 0.9 | 83.41 | 82.23 |
| 49 | 2,2',4,5'-Tetrachlorobiphenyl | 0.098 | 87.4± 0.8 | 89.48 | 87.73 |
| 53 | 2,2',5,6'-Tetrachlorobiphenyl | 0.094 | 84.9± 0.6 | 89.48 | 85.11 |
| 95 | 2,2',3,5',6-Pentachlorobiphenyl | 0.103 | 92.3± 0.6 | 95.54 | 93.89 |
| 96 | 2,2',3,6,6'-Pentachlorobiphenyl | 0.100 | 89.6± 0.2 | 95.54 | 91.92 |
| 103 | 2,2',4,5',6-Pentachlorobiphenyl | 0.105 | 91.6± 0.5 | 95.54 | 95.20 |
| 153 | 2,2',4,4',5,5'-Hexachlorobiphenyl | 0.110 | 103.5± 0.1 | 101.60 | 101.40 |
| 156 | 2,3,3',4,4',5-Hexachlorobiphenyl | 0.119 | 112.6± 0.4 | 101.60 | 107.20 |
| 171 | 2,2',3,3',4,4',6-Heptachlorobiphenyl | 0.116 | 109.1± 0.0 | 107.70 | 108.20 |

* Experimental values taken from Reference [14].
[a] Calculated $\Delta_{vap}H_m$ using one parameter ($N_{NH}$) regression.
[b] Calculated $\Delta_{vap}H_m$ using two parameter ($\omega$, $N_{NH}$) regression.

**Table 3**. Enthalpy of vaporization for the data set of 27 Polychlorinated Biphenyls

| IUPAC No | Compounds | $\omega$ | ($\Delta_{vap}H_m$ (298.15 K)) | | |
|---|---|---|---|---|---|
| | | | Exptl * | Calcd[a] ($N_{NH}$) | Calcd[b] ($\omega$, $N_{NH}$) |
| | Biphenyl | 0.077 | 58.36±3.58 | 62.83 | 62.32 |
| 4 | 2,2'-Dichlorobiphenyl | 0.086 | 71.27±1.09 | 72.90 | 72.24 |
| 8 | 2,4'-Dichlorobiphenyl | 0.087 | 75.75±1.47 | 72.90 | 72.37 |
| 15 | 4,4'-Dichlorobiphenyl | 0.090 | 79.17±1.15 | 72.90 | 72.75 |
| 17 | 2,2',4-Trichlorobiphenyl | 0.095 | 73.45±1.69 | 77.94 | 77.77 |
| 18 | 2,2',5-Trichlorobiphenyl | 0.094 | 76.25±1.18 | 77.94 | 77.64 |
| 22 | 2,3,4'-Trichlorobiphenyl | 0.100 | 76.45±3.38 | 77.94 | 78.41 |
| 28 | 2,4,4'-Trichlorobiphenyl | 0.102 | 79.44±0.86 | 77.94 | 78.67 |
| 31 | 2,4',5-Trichlorobiphenyl | 0.097 | 83.57±1.40 | 77.94 | 78.03 |
| 33 | 2,3',4'-Trichlorobiphenyl | 0.100 | 81.55±1.94 | 77.94 | 78.41 |
| 44 | 2,2',3,5'-Tetrachlorobiphenyl | 0.097 | 83.26±4.26 | 82.98 | 82.41 |
| 52 | 2,2',5,5'-Tetrachlorobiphenyl | 0.097 | 81.28±0.47 | 82.98 | 82.41 |
| 66 | 2,3',4,4'-Tetrachlorobiphenyl | 0.108 | 78±2.59 | 82.98 | 83.82 |
| 70 | 2,3',4',5-Tetrachlorobiphenyl | 0.107 | 83.41±2.28 | 82.98 | 83.69 |
| 77 | 3,3',4,4'-Tetrachlorobiphenyl | 0.112 | 83.47±1.85 | 82.98 | 84.33 |
| 95 | 2,2',3,5',6-Pentachlorobiphenyl | 0.103 | 82.95±4.05 | 88.02 | 87.56 |
| 98 | 2,2',3,4',6'-Pentachlorobiphenyl | 0.105 | 82.81±4.22 | 88.02 | 87.82 |
| 101 | 2,2',4,5,5'-Pentachlorobiphenyl | 0.106 | 87.11±4.64 | 88.02 | 87.95 |
| 110 | 2,3,3',4',6-Pentachlorobiphenyl | 0.106 | 89.58±1.45 | 88.02 | 87.95 |
| 118 | 2,3',4,4',5-Pentachlorobiphenyl | 0.114 | 87.03±3.88 | 88.02 | 88.97 |
| 126 | 3,3',4,4',5-Pentachlorobiphenyl | 0.108 | 95.38±2.92 | 88.02 | 88.20 |
| 137 | 2,2',3,4,4',5-Hexachlorobiphenyl | 0.113 | 94.46±0.88 | 93.06 | 93.22 |
| 149 | 2,2',3,4',5',6-Hexachlorobiphenyl | 0.107 | 90.01±3.13 | 93.06 | 92.46 |
| 153 | 2,2',4,4',5,5'-Hexachlorobiphenyl | 0.110 | 99.38±4.16 | 93.06 | 92.84 |
| 170 | 2,2', 3,3', 4,4',5-Heptachlorobiphenyl | 0.116 | 97.38±2.15 | 98.10 | 97.99 |
| 174 | 2,2', 3,3', 4,5,6'-Heptachlorobiphenyl | 0.113 | 95.56±1.79 | 98.10 | 97.61 |
| 180 | 2,2', 3,4,4', 5,5'-Heptachlorobiphenyl | 0.115 | 99.37±2.54 | 98.10 | 97.86 |

*Experimental values taken from reference [15].
[a]Calculated $\Delta_{vap}H_m$ using one parameter ($N_{NH}$) regression.
[b]Calculated $\Delta_{vap}H_m$ using two parameter ($\omega$, $N_{NH}$) regression.

**Table 4**. Predicted ($\Delta_{vap}H_m$ (298.15 K)) for 193 PCBs using regression from 17 PCBs

| IUPAC No | Compounds | $\omega$ | ($\Delta_{vap}H_m$ (298.15 K) Calcd[a] ($N_{NH}$) | Calcd[b] ($\omega$, $N_{NH}$) | $\Delta_{vap}H_m$[c] |
|---|---|---|---|---|---|
| 4 | 2,2'-Dichlorobiphenyl | 0.086 | 77.34 | 74.11 | 72.6 |
| 5 | 2,3-Dichlorobiphenyl | 0.094 | 77.34 | 79.35 | 73.4 |
| 6 | 2,3'-Dichlorobiphenyl | 0.094 | 77.34 | 79.35 | 77.8 |
| 8 | 2,4'-Dichlorobiphenyl | 0.087 | 77.34 | 74.76 | 76.5 |
| 10 | 2,6-Dichlorobiphenyl | 0.091 | 77.34 | 77.38 | 80.6 |
| 12 | 3,4-Dichlorobiphenyl | 0.099 | 77.34 | 82.62 | 79.5 |
| 13 | 3,4'-Dichlorobiphenyl | 0.099 | 77.34 | 82.62 | 82 |
| 14 | 3,5-Dichlorobiphenyl | 0.099 | 77.34 | 82.62 | 81.4 |
| 16 | 2,2',3-Trichlorobiphenyl | 0.092 | 83.41 | 80.92 | 79.9 |
| 17 | 2,2',4-Trichlorobiphenyl | 0.095 | 83.41 | 82.88 | 81.6 |
| 19 | 2,2',6-Trichlorobiphenyl | 0.092 | 83.41 | 80.92 | 81 |
| 20 | 2,3,3'-Trichlorobiphenyl | 0.100 | 83.41 | 86.16 | 93.7 |
| 21 | 2,3,4-Trichlorobiphenyl | 0.102 | 83.41 | 87.47 | 88.6 |
| 22 | 2,3,4'-Trichlorobiphenyl | 0.100 | 83.41 | 86.16 | 81 |
| 23 | 2,3,5-Trichlorobiphenyl | 0.102 | 83.41 | 87.47 | 90.1 |
| 24 | 2,3,6-Trichlorobiphenyl | 0.100 | 83.41 | 86.16 | 89.2 |
| 25 | 2,3',4-Trichlorobiphenyl | 0.102 | 83.41 | 87.47 | 93.2 |
| 26 | 2,3',5-Trichlorobiphenyl | 0.101 | 83.41 | 86.81 | 89.4 |
| 27 | 2,3',6-Trichlorobiphenyl | 0.094 | 83.41 | 82.23 | 80.7 |
| 28 | 2,4,4'-Trichlorobiphenyl | 0.102 | 83.41 | 87.47 | 89.3 |
| 29 | 2,4,5-Trichlorobiphenyl | 0.103 | 83.41 | 88.12 | 82.9 |
| 30 | 2,4,6-Trichlorobiphenyl | 0.102 | 83.41 | 87.47 | 91.1 |
| 31 | 2,4',5-Trichlorobiphenyl | 0.097 | 83.41 | 84.19 | 91.2 |
| 32 | 2,4',6-Trichlorobiphenyl | 0.094 | 83.41 | 82.23 | 82.9 |
| 33 | 2,3',4'-Trichlorobiphenyl | 0.100 | 83.41 | 86.16 | 88.3 |
| 34 | 2,3',5'-Trichlorobiphenyl | 0.093 | 83.41 | 81.57 | 83.3 |
| 35 | 3,3',4-Trichlorobiphenyl | 0.106 | 83.41 | 90.09 | 90.1 |
| 36 | 3,3',5-Trichlorobiphenyl | 0.096 | 83.41 | 83.54 | 85.7 |
| 37 | 3,4,4'-Trichlorobiphenyl | 0.106 | 83.41 | 90.09 | 91.1 |
| 38 | 3,4,5-Trichlorobiphenyl | 0.106 | 83.41 | 90.09 | 89.2 |
| 39 | 3,4',5-Trichlorobiphenyl | 0.106 | 83.41 | 90.09 | 86.8 |
| 40 | 2,2',3,3'-Tetrachlorobiphenyl | 0.096 | 89.48 | 86.42 | 94.7 |
| 41 | 2,2',3,4-Tetrachlorobiphenyl | 0.101 | 89.48 | 89.69 | 90.5 |
| 42 | 2,2',3,4'-Tetrachlorobiphenyl | 0.099 | 89.48 | 88.38 | 95.2 |
| 43 | 2,2',3,5-Tetrachlorobiphenyl | 0.101 | 89.48 | 89.69 | 86.3 |
| 44 | 2,2',3,5'-Tetrachlorobiphenyl | 0.097 | 89.48 | 87.07 | 84.4 |
| 45 | 2,2',3,6-Tetrachlorobiphenyl | 0.100 | 89.48 | 89.04 | 80.5 |
| 46 | 2,2',3,6'-Tetrachlorobiphenyl | 0.095 | 89.48 | 85.76 | 85.9 |
| 47 | 2,2',4,4'-Tetrachlorobiphenyl | 0.100 | 89.48 | 89.04 | 90.3 |
| 48 | 2,2',4,5-Tetrachlorobiphenyl | 0.102 | 89.48 | 90.35 | 90.8 |
| 50 | 2,2',4,6-Tetrachlorobiphenyl | 0.102 | 89.48 | 90.35 | 81.2 |
| 51 | 2,2',4,6'-Tetrachlorobiphenyl | 0.096 | 89.48 | 86.42 | 86.5 |
| 52 | 2,2',5,5'-Tetrachlorobiphenyl | 0.097 | 89.48 | 87.07 | 84.5 |
| 54 | 2,2',6,6'-Tetrachlorobiphenyl | 0.093 | 89.48 | 84.45 | 79.4 |
| 55 | 2,3,3',4-Tetrachlorobiphenyl | 0.107 | 89.48 | 93.62 | 102.8 |
| 56 | 2,3,3',4'-Tetrachlorobiphenyl | 0.105 | 89.48 | 92.31 | 99.7 |
| 57 | 2,3,3',5-Tetrachlorobiphenyl | 0.107 | 89.48 | 93.62 | 96.1 |
| 58 | 2,3,3',5'-Tetrachlorobiphenyl | 0.098 | 89.48 | 87.73 | 101.9 |

| 59 | 2,3,3',6-Tetrachlorobiphenyl | 0.104 | 89.48 | 91.66 | 89.8 |
|---|---|---|---|---|---|
| 60 | 2,3,4,4'-Tetrachlorobiphenyl | 0.107 | 89.48 | 93.62 | 98.6 |
| 61 | 2,3,4,5-Tetrachlorobiphenyl | 0.109 | 89.48 | 94.93 | 93.8 |
| 62 | 2,3,4,6-Tetrachlorobiphenyl | 0.108 | 89.48 | 94.28 | 97.6 |
| 63 | 2,3,4',5-Tetrachlorobiphenyl | 0.107 | 89.48 | 93.62 | 94.1 |
| 64 | 2,3,4',6-Tetrachlorobiphenyl | 0.104 | 89.48 | 91.66 | 91.7 |
| 65 | 2,3,5,6-Tetrachlorobiphenyl | 0.108 | 89.48 | 94.28 | 97.8 |
| 66 | 2,3',4,4'-Tetrachlorobiphenyl | 0.108 | 89.48 | 94.28 | 93 |
| 67 | 2,3',4,5-Tetrachlorobiphenyl | 0.109 | 89.48 | 94.93 | 96.6 |
| 68 | 2,3',4,5'-Tetrachlorobiphenyl | 0.102 | 89.48 | 90.35 | 102.5 |
| 69 | 2,3',4,6-Tetrachlorobiphenyl | 0.105 | 89.48 | 92.31 | 90.3 |
| 70 | 2,3',4',5-Tetrachlorobiphenyl | 0.107 | 89.48 | 93.62 | 96.1 |
| 71 | 2,3',4',6-Tetrachlorobiphenyl | 0.097 | 89.48 | 87.07 | 92.8 |
| 72 | 2,3',5,5'-Tetrachlorobiphenyl | 0.101 | 89.48 | 89.69 | 97.8 |
| 73 | 2,3',5',6-Tetrachlorobiphenyl | 0.098 | 89.48 | 87.73 | 90.2 |
| 74 | 2,4,4',5-Tetrachlorobiphenyl | 0.106 | 89.48 | 92.97 | 94.4 |
| 75 | 2,4,4',6-Tetrachlorobiphenyl | 0.104 | 89.48 | 91.66 | 92.4 |
| 76 | 2,3',4',5'-Tetrachlorobiphenyl | 0.102 | 89.48 | 90.35 | 92.4 |
| 77 | 3,3',4,4'-Tetrachlorobiphenyl | 0.112 | 89.48 | 96.90 | 97.8 |
| 78 | 3,3',4,5-Tetrachlorobiphenyl | 0.105 | 89.48 | 92.31 | 94.8 |
| 79 | 3,3',4,5'-Tetrachlorobiphenyl | 0.101 | 89.48 | 89.69 | 99.7 |
| 80 | 3,3',5,5'-Tetrachlorobiphenyl | 0.100 | 89.48 | 89.04 | 95.3 |
| 81 | 3,4,4',5-Tetrachlorobiphenyl | 0.112 | 89.48 | 96.90 | 95.8 |
| 82 | 2,2',3,3',4-Pentachlorobiphenyl | 0.105 | 95.54 | 95.20 | 92.5 |
| 83 | 2,2',3,3',5-Pentachlorobiphenyl | 0.104 | 95.54 | 94.54 | 91.7 |
| 84 | 2,2',3,3',6-Pentachlorobiphenyl | 0.103 | 95.54 | 93.89 | 89.2 |
| 85 | 2,2',3,4,4'-Pentachlorobiphenyl | 0.105 | 95.54 | 95.20 | 100.2 |
| 86 | 2,2', 3,4,5-Pentachlorobiphenyl | 0.109 | 95.54 | 97.82 | 93.1 |
| 87 | 2,2', 3,4,5'-Pentachlorobiphenyl | 0.104 | 95.54 | 94.54 | 99.6 |
| 88 | 2,2', 3,4,6-Pentachlorobiphenyl | 0.109 | 95.54 | 97.82 | 89 |
| 89 | 2,2',3,4,6'-Pentachlorobiphenyl | 0.102 | 95.54 | 93.23 | 88.6 |
| 90 | 2,2',3,4',5-Pentachlorobiphenyl | 0.105 | 95.54 | 95.20 | 96 |
| 91 | 2,2',3,4',6-Pentachlorobiphenyl | 0.104 | 95.54 | 94.54 | 89.2 |
| 92 | 2,2',3,5,5'-Pentachlorobiphenyl | 0.104 | 95.54 | 94.54 | 97 |
| 93 | 2,2',3,5,6-Pentachlorobiphenyl | 0.108 | 95.54 | 97.16 | 88.5 |
| 94 | 2,2',3,5,6'-Pentachlorobiphenyl | 0.101 | 95.54 | 92.58 | 88.1 |
| 97 | 2,2',3,4',5'-Pentachlorobiphenyl | 0.106 | 95.54 | 95.85 | 96.3 |
| 98 | 2,2',3,4',6'-Pentachlorobiphenyl | 0.105 | 95.54 | 95.20 | 96.3 |
| 99 | 2,2',4,4',5-Pentachlorobiphenyl | 0.107 | 95.54 | 96.51 | 102.3 |
| 100 | 2,2',4,4',6-Pentachlorobiphenyl | 0.106 | 95.54 | 95.85 | 92.1 |
| 101 | 2,2',4,5,5'-Pentachlorobiphenyl | 0.106 | 95.54 | 95.85 | 95.6 |
| 102 | 2,2',4,5,6'-Pentachlorobiphenyl | 0.103 | 95.54 | 93.89 | 88.4 |
| 104 | 2,2',4,6,6'-Pentachlorobiphenyl | 0.103 | 95.54 | 93.89 | 86.3 |
| 105 | 2,3,3',4,4'-Pentachlorobiphenyl | 0.112 | 95.54 | 99.78 | 106.7 |
| 106 | 2,3,3',4,5-Pentachlorobiphenyl | 0.114 | 95.54 | 101.10 | 106.9 |
| 107 | 2,3,3',4',5-Pentachlorobiphenyl | 0.112 | 95.54 | 99.78 | 102.2 |
| 108 | 2,3,3',4,5'-Pentachlorobiphenyl | 0.108 | 95.54 | 97.16 | 100.9 |
| 109 | 2,3,3',4,6-Pentachlorobiphenyl | 0.112 | 95.54 | 99.78 | 95.2 |
| 110 | 2,3,3',4',6-Pentachlorobiphenyl | 0.106 | 95.54 | 95.85 | 97.6 |
| 111 | 2,3,3',5,5'-Pentachlorobiphenyl | 0.108 | 95.54 | 97.16 | 102 |
| 112 | 2,3,3',5,6-Pentachlorobiphenyl | 0.112 | 95.54 | 99.78 | 94.7 |
| 113 | 2,3,3',5',6-Pentachlorobiphenyl | 0.107 | 95.54 | 96.51 | 95.1 |
| 114 | 2,3,4,4',5-Pentachlorobiphenyl | 0.112 | 95.54 | 99.78 | 107.1 |

| 115 | 2,3,4,4',6-Pentachlorobiphenyl | 0.112 | 95.54 | 99.78 | 97.4 |
| --- | --- | --- | --- | --- | --- |
| 116 | 2,3,4,5,6-Pentachlorobiphenyl | 0.114 | 95.54 | 101.10 | 108 |
| 117 | 2,3,4',5,6-Pentachlorobiphenyl | 0.112 | 95.54 | 99.78 | 96.9 |
| 118 | 2,3',4,4',5-Pentachlorobiphenyl | 0.114 | 95.54 | 101.10 | 107.6 |
| 119 | 2,3',4,4',6-Pentachlorobiphenyl | 0.108 | 95.54 | 97.16 | 102.2 |
| 120 | 2,3',4,5,5'-Pentachlorobiphenyl | 0.110 | 95.54 | 98.47 | 106.2 |
| 121 | 2,3',4,5',6-Pentachlorobiphenyl | 0.109 | 95.54 | 97.82 | 99.7 |
| 122 | 2,3,3',4',5'-Pentachlorobiphenyl | 0.106 | 95.54 | 95.85 | 103.7 |
| 123 | 2,3',4,4',5'-Pentachlorobiphenyl | 0.106 | 95.54 | 95.85 | 104.4 |
| 124 | 2,3',4',5,5'-Pentachlorobiphenyl | 0.105 | 95.54 | 95.20 | 104.6 |
| 125 | 2,3',4',5',6-Pentachlorobiphenyl | 0.103 | 95.54 | 93.89 | 96 |
| 126 | 3,3',4,4',5-Pentachlorobiphenyl | 0.108 | 95.54 | 97.16 | 107.4 |
| 127 | 3,3',4,5,5'-Pentachlorobiphenyl | 0.109 | 95.54 | 97.82 | 104.4 |
| 128 | 2,2',3,3',4,4'-Hexachlorobiphenyl | 0.109 | 101.60 | 100.70 | 106 |
| 129 | 2,2',3,3',4,5-Hexachlorobiphenyl | 0.112 | 101.60 | 102.70 | 99.3 |
| 130 | 2,2',3,3',4,5'-Hexachlorobiphenyl | 0.108 | 101.60 | 100.00 | 113.2 |
| 131 | 2,2',3,3',4,6-Hexachlorobiphenyl | 0.112 | 101.60 | 102.70 | 96.6 |
| 132 | 2,2',3,3',4,6'-Hexachlorobiphenyl | 0.107 | 101.60 | 99.39 | 101.1 |
| 133 | 2,2',3,3',5,5'-Hexachlorobiphenyl | 0.108 | 101.60 | 100.00 | 96.8 |
| 134 | 2,2',3,3',5,6-Hexachlorobiphenyl | 0.111 | 101.60 | 102.00 | 92.1 |
| 135 | 2,2',3,3',5,6'-Hexachlorobiphenyl | 0.106 | 101.60 | 98.73 | 97 |
| 136 | 2,2',3,3',6,6'-Hexachlorobiphenyl | 0.103 | 101.60 | 96.77 | 94.2 |
| 137 | 2,2',3,4,4',5-Hexachlorobiphenyl | 0.113 | 101.60 | 103.30 | 105.3 |
| 138 | 2,2',3,4,4',5'-Hexachlorobiphenyl | 0.109 | 101.60 | 100.70 | 103.1 |
| 139 | 2,2',3,4,4',6-Hexachlorobiphenyl | 0.113 | 101.60 | 103.30 | 102.5 |
| 140 | 2,2',3,4,4',6'-Hexachlorobiphenyl | 0.109 | 101.60 | 100.70 | 100.9 |
| 141 | 2,2',3,4,5,5'-Hexachlorobiphenyl | 0.112 | 101.60 | 102.70 | 106.2 |
| 142 | 2,2',3,4,5,6-Hexachlorobiphenyl | 0.114 | 101.60 | 104.00 | 91 |
| 143 | 2,2',3,4,5,6'-Hexachlorobiphenyl | 0.109 | 101.60 | 100.70 | 94.6 |
| 144 | 2,2',3,4,5',6-Hexachlorobiphenyl | 0.112 | 101.60 | 102.70 | 96.5 |
| 145 | 2,2',3,4,6,6'-Hexachlorobiphenyl | 0.109 | 101.60 | 100.70 | 92.3 |
| 146 | 2,2',3,4',5,5'-Hexachlorobiphenyl | 0.110 | 101.60 | 101.40 | 106.8 |
| 147 | 2,2',3,4',5,6-Hexachlorobiphenyl | 0.112 | 101.60 | 102.70 | 98.3 |
| 148 | 2,2',3,4',5,6'-Hexachlorobiphenyl | 0.108 | 101.60 | 100.00 | 96.5 |
| 149 | 2,2',3,4',5',6-Hexachlorobiphenyl | 0.107 | 101.60 | 99.39 | 97.1 |
| 150 | 2,2',3,4',6,6'-Hexachlorobiphenyl | 0.106 | 101.60 | 98.73 | 95.7 |
| 151 | 2,2',3,5,5',6-Hexachlorobiphenyl | 0.111 | 101.60 | 102.00 | 92 |
| 152 | 2,2',3,5,6,6'-Hexachlorobiphenyl | 0.108 | 101.60 | 100.00 | 87.8 |
| 154 | 2,2',4,4',5,6'-Hexachlorobiphenyl | 0.108 | 101.60 | 100.00 | 96.9 |
| 155 | 2,2',4,4',6,6'-Hexachlorobiphenyl | 0.108 | 101.60 | 100.00 | 96.3 |
| 157 | 2,3,3',4,4',5'-Hexachlorobiphenyl | 0.111 | 101.60 | 102.00 | 117 |
| 158 | 2,3,3',4,4',6-Hexachlorobiphenyl | 0.115 | 101.60 | 104.60 | 110.7 |
| 159 | 2,3,3',4,5,5'-Hexachlorobiphenyl | 0.116 | 101.60 | 105.30 | 111.4 |
| 160 | 2,3,3',4,5,6-Hexachlorobiphenyl | 0.118 | 101.60 | 106.60 | 103.5 |
| 161 | 2,3,3',4,5',6-Hexachlorobiphenyl | 0.116 | 101.60 | 105.30 | 108.4 |
| 162 | 2,3,3',4',5,5'-Hexachlorobiphenyl | 0.111 | 101.60 | 102.00 | 112.6 |
| 163 | 2,3,3',4',5,6-Hexachlorobiphenyl | 0.115 | 101.60 | 104.60 | 106.7 |
| 164 | 2,3,3',4',5',6-Hexachlorobiphenyl | 0.110 | 101.60 | 101.40 | 111.3 |
| 165 | 2,3,3',5,5',6-Hexachlorobiphenyl | 0.115 | 101.60 | 104.60 | 104.2 |
| 166 | 2,3,4,4',5,6-Hexachlorobiphenyl | 0.118 | 101.60 | 106.60 | 106 |
| 167 | 2,3',4,4',5,5'-Hexachlorobiphenyl | 0.112 | 101.60 | 102.70 | 112.2 |
| 168 | 2,3',4,4',5',6-Hexachlorobiphenyl | 0.112 | 101.60 | 102.70 | 112.2 |
| 169 | 3,3',4,4',5,5'-Hexachlorobiphenyl | 0.112 | 101.60 | 102.70 | 112.2 |
| 170 | 2,2',3,3',4,4',5-Heptachlorobiphenyl | 0.116 | 107.70 | 108.20 | 116 |
| 172 | 2,2',3,3',4,5,5'-Heptachlorobiphenyl | 0.116 | 107.70 | 108.20 | 117.2 |
| 173 | 2,2',3,3',4,5,6-Heptachlorobiphenyl | 0.118 | 107.70 | 109.50 | 99.6 |
| 174 | 2,2',3,3',4,5,6'-Heptachlorobiphenyl | 0.113 | 107.70 | 106.20 | 103.3 |
| 175 | 2,2',3,3',4,5',6-Heptachlorobiphenyl | 0.115 | 107.70 | 107.50 | 114.3 |
| 176 | 2,2',3,3',4,6,6'-Heptachlorobiphenyl | 0.112 | 107.70 | 105.50 | 100.8 |
| 177 | 2,2',3,3',4,5,6'-Heptachlorobiphenyl | 0.114 | 107.70 | 106.90 | 102.9 |
| 178 | 2,2',3,3',5,5',6-Heptachlorobiphenyl | 0.114 | 107.70 | 106.90 | 105.5 |
| 179 | 2,2',3,3',5,6,6'-Heptachlorobiphenyl | 0.111 | 107.70 | 104.90 | 96.3 |
| 180 | 2,2',3,4,4',5,5'-Heptachlorobiphenyl | 0.115 | 107.70 | 107.50 | 111.9 |
| 181 | 2,2',3,4,4',5,6-Heptachlorobiphenyl | 0.119 | 107.70 | 110.10 | 102 |
| 182 | 2,2',3,4,4',5,6'-Heptachlorobiphenyl | 0.113 | 107.70 | 106.20 | 102.8 |
| 183 | 2,2',3,4,4',5',6-Heptachlorobiphenyl | 0.115 | 107.70 | 107.50 | 113.1 |
| 184 | 2,2',3,4,4',6,6'-Heptachlorobiphenyl | 0.113 | 107.70 | 106.20 | 102.2 |
| 185 | 2,2',3,4,5,5',6-Heptachlorobiphenyl | 0.118 | 107.70 | 109.50 | 99.6 |
| 186 | 2,2',3,4,5,6,6'-Heptachlorobiphenyl | 0.115 | 107.70 | 107.50 | 94.5 |
| 187 | 2,2',3,4',5,5',6-Heptachlorobiphenyl | 0.114 | 107.70 | 106.90 | 106 |
| 188 | 2,2',3,4',5,6,6'-Heptachlorobiphenyl | 0.111 | 107.70 | 104.90 | 97.9 |
| 189 | 2,3,3',4,4',5,5'-Heptachlorobiphenyl | 0.119 | 107.70 | 110.10 | 122.1 |
| 190 | 2,3,3',4,4',5,6-Heptachlorobiphenyl | 0.121 | 107.70 | 111.40 | 115.7 |
| 191 | 2,3,3',4,4',5',6-Heptachlorobiphenyl | 0.119 | 107.70 | 110.10 | 120.5 |
| 192 | 2,3,3',4,5,5',6-Heptachlorobiphenyl | 0.122 | 107.70 | 112.10 | 113.5 |
| 193 | 2,3,3',4',5,5',6-Heptachlorobiphenyl | 0.118 | 107.70 | 109.50 | 113.2 |
| 194 | 2,2',3,3',4,4',5,5'-Octachlorobiphenyl | 0.119 | 113.70 | 113.00 | 119.6 |
| 195 | 2,2',3,3',4,4',5,6-Octachlorobiphenyl | 0.121 | 113.70 | 114.30 | 117.3 |
| 196 | 2,2',3,3',4,4',5,6'-Octachlorobiphenyl | 0.118 | 113.70 | 112.40 | 111.4 |
| 197 | 2,2',3,3',4,4',6,6'-Octachlorobiphenyl | 0.116 | 113.70 | 111.00 | 110.5 |
| 198 | 2,2',3,3',4,5,5',6-Octachlorobiphenyl | 0.121 | 113.70 | 114.30 | 108.1 |
| 199 | 2,2',3,3',4,5,5',6'-Octachlorobiphenyl | 0.117 | 113.70 | 111.70 | 112.1 |
| 200 | 2,2',3,3',4,5,6,6'-Octachlorobiphenyl | 0.118 | 113.70 | 112.40 | 102.8 |
| 201 | 2,2',3,3',4,5',6,6'-Octachlorobiphenyl | 0.115 | 113.70 | 110.40 | 109.2 |
| 202 | 2,2',3,3',5,5',6,6'-Octachlorobiphenyl | 0.114 | 113.70 | 109.70 | 104.7 |
| 203 | 2,2',3,4,4',5,5',6-Octachlorobiphenyl | 0.121 | 113.70 | 114.30 | 110.7 |
| 204 | 2,2',3,4,4',5,6,6'-Octachlorobiphenyl | 0.119 | 113.70 | 113.00 | 104.4 |
| 205 | 2,3,3',4,4',5,5',6-Octachlorobiphenyl | 0.125 | 113.70 | 116.90 | 125.5 |
| 206 | 2,2',3,3',4,4',5,5',6-Nonachlorobiphenyl | 0.124 | 119.80 | 119.20 | 119 |
| 207 | 2,2',3,3',4,4',5,6,6'-Nonachlorobiphenyl | 0.122 | 119.80 | 117.90 | 115.4 |
| 208 | 2,2',3,3',4,5,5',6,6'-Nonachlorobiphenyl | 0.121 | 119.80 | 117.20 | 114.9 |
| 209 | 2,2',3,3',4,4',5,5',6,6'-Decachlorobiphenyl | 0.124 | 119.80 | 119.20 | 121.4 |

[a]Calculated $\Delta_{vap}H_m$ using one parameter ($N_{NH}$) regression.
[b]Calculated $\Delta_{vap}H_m$ using two parameter ($\omega$, $N_{NH}$) regression.
[c]Data from [Ref.16] for comparison.

**Table 5:** Predicted ($\Delta_{vap}H_m$ (298.15 K)) for 183 PCBs using regression from 27 PCBs

| IUPAC No | Compounds | $\omega$ | ($\Delta_{vap}H_m$ (298.15 K)) Calcd[a] ($N_{NH}$) | Calcd[b] ($\omega$, $N_{NH}$) |
|---|---|---|---|---|
| 1 | 2-Chlorobiphenyl | 0.087 | 67.87 | 66.81 |
| 2 | 3-Chlorobiphenyl | 0.085 | 67.87 | 66.55 |
| 3 | 4-Chlorobiphenyl | 0.085 | 67.87 | 66.55 |
| 5 | 2,3-Dichlorobiphenyl | 0.094 | 72.90 | 72.09 |
| 6 | 2,3'-Dichlorobiphenyl | 0.094 | 72.90 | 72.09 |
| 7 | 2,4-Dichlorobiphenyl | 0.096 | 72.90 | 72.34 |
| 9 | 2,5-Dichlorobiphenyl | 0.095 | 72.90 | 72.22 |
| 10 | 2,6-Dichlorobiphenyl | 0.091 | 72.90 | 71.70 |
| 11 | 3,3'-Dichlorobiphenyl | 0.099 | 72.90 | 72.73 |
| 12 | 3,4-Dichlorobiphenyl | 0.099 | 72.90 | 72.73 |
| 13 | 3,4'-Dichlorobiphenyl | 0.099 | 72.90 | 72.73 |
| 14 | 3,5-Dichlorobiphenyl | 0.099 | 72.90 | 72.73 |
| 16 | 2,2',3-Trichlorobiphenyl | 0.092 | 77.94 | 76.22 |
| 19 | 2,2',6-Trichlorobiphenyl | 0.092 | 77.94 | 76.22 |
| 20 | 2,3,3'-Trichlorobiphenyl | 0.100 | 77.94 | 77.24 |
| 21 | 2,3,4-Trichlorobiphenyl | 0.102 | 77.94 | 77.50 |
| 23 | 2,3,5-Trichlorobiphenyl | 0.102 | 77.94 | 77.50 |
| 24 | 2,3,6-Trichlorobiphenyl | 0.100 | 77.94 | 77.24 |
| 25 | 2,3',4-Trichlorobiphenyl | 0.102 | 77.94 | 77.50 |
| 26 | 2,3',5-Trichlorobiphenyl | 0.101 | 77.94 | 77.37 |
| 27 | 2,3',6-Trichlorobiphenyl | 0.094 | 77.94 | 76.47 |
| 29 | 2,4,5-Trichlorobiphenyl | 0.103 | 77.94 | 77.62 |
| 30 | 2,4,6-Trichlorobiphenyl | 0.102 | 77.94 | 77.50 |
| 32 | 2,4',6-Trichlorobiphenyl | 0.094 | 77.94 | 76.47 |
| 34 | 2,3',5'-Trichlorobiphenyl | 0.093 | 77.94 | 76.34 |
| 35 | 3,3',4-Trichlorobiphenyl | 0.106 | 77.94 | 78.01 |
| 36 | 3,3',5-Trichlorobiphenyl | 0.096 | 77.94 | 76.73 |
| 37 | 3,4,4'-Trichlorobiphenyl | 0.106 | 77.94 | 78.01 |
| 38 | 3,4,5-Trichlorobiphenyl | 0.106 | 77.94 | 78.01 |
| 39 | 3,4',5-Trichlorobiphenyl | 0.106 | 77.94 | 78.01 |
| 40 | 2,2',3,3'-Tetrachlorobiphenyl | 0.096 | 82.98 | 81.11 |
| 41 | 2,2',3,4-Tetrachlorobiphenyl | 0.101 | 82.98 | 81.75 |
| 42 | 2,2',3,4'-Tetrachlorobiphenyl | 0.099 | 82.98 | 81.49 |
| 43 | 2,2',3,5-Tetrachlorobiphenyl | 0.101 | 82.98 | 81.75 |
| 45 | 2,2',3,6-Tetrachlorobiphenyl | 0.100 | 82.98 | 81.62 |
| 46 | 2,2',3,6'-Tetrachlorobiphenyl | 0.095 | 82.98 | 80.98 |
| 47 | 2,2',4,4'-Tetrachlorobiphenyl | 0.100 | 82.98 | 81.62 |
| 48 | 2,2',4,5-Tetrachlorobiphenyl | 0.102 | 82.98 | 81.88 |
| 49 | 2,2',4,5'-Tetrachlorobiphenyl | 0.098 | 82.98 | 81.37 |
| 50 | 2,2',4,6-Tetrachlorobiphenyl | 0.102 | 82.98 | 81.88 |
| 51 | 2,2',4,6'-Tetrachlorobiphenyl | 0.096 | 82.98 | 81.11 |
| 53 | 2,2',5,6'-Tetrachlorobiphenyl | 0.094 | 82.98 | 80.86 |
| 54 | 2,2',6,6'-Tetrachlorobiphenyl | 0.093 | 82.98 | 80.73 |
| 55 | 2,3,3',4-Tetrachlorobiphenyl | 0.107 | 82.98 | 82.52 |
| 56 | 2,3,3',4'-Tetrachlorobiphenyl | 0.105 | 82.98 | 82.26 |
| 57 | 2,3,3',5-Tetrachlorobiphenyl | 0.107 | 82.98 | 82.52 |
| 58 | 2,3,3',5'-Tetrachlorobiphenyl | 0.098 | 82.98 | 81.37 |
| 59 | 2,3,3',6-Tetrachlorobiphenyl | 0.104 | 82.98 | 82.13 |
| 60 | 2,3,4,4'-Tetrachlorobiphenyl | 0.107 | 82.98 | 82.52 |
| 61 | 2,3,4,5-Tetrachlorobiphenyl | 0.109 | 82.98 | 82.77 |
| 62 | 2,3,4,6-Tetrachlorobiphenyl | 0.108 | 82.98 | 82.65 |
| 63 | 2,3,4',5-Tetrachlorobiphenyl | 0.107 | 82.98 | 82.52 |
| 64 | 2,3,4',6-Tetrachlorobiphenyl | 0.104 | 82.98 | 82.13 |
| 65 | 2,3,5,6-Tetrachlorobiphenyl | 0.108 | 82.98 | 82.65 |
| 67 | 2,3',4,5-Tetrachlorobiphenyl | 0.109 | 82.98 | 82.77 |
| 68 | 2,3',4,5'-Tetrachlorobiphenyl | 0.102 | 82.98 | 81.88 |
| 69 | 2,3',4,6-Tetrachlorobiphenyl | 0.105 | 82.98 | 82.26 |
| 71 | 2,3',4',6-Tetrachlorobiphenyl | 0.097 | 82.98 | 81.24 |
| 72 | 2,3',5,5'-Tetrachlorobiphenyl | 0.101 | 82.98 | 81.75 |
| 73 | 2,3',5',6-Tetrachlorobiphenyl | 0.098 | 82.98 | 81.37 |
| 74 | 2,4,4',5-Tetrachlorobiphenyl | 0.106 | 82.98 | 82.39 |
| 75 | 2,4,4',6-Tetrachlorobiphenyl | 0.104 | 82.98 | 82.13 |
| 76 | 2,3',4',5'-Tetrachlorobiphenyl | 0.102 | 82.98 | 81.88 |
| 78 | 3,3',4,5-Tetrachlorobiphenyl | 0.105 | 82.98 | 82.26 |
| 79 | 3,3',4,5'-Tetrachlorobiphenyl | 0.101 | 82.98 | 81.75 |
| 80 | 3,3',5,5'-Tetrachlorobiphenyl | 0.100 | 82.98 | 81.62 |
| 81 | 3,4,4',5-Tetrachlorobiphenyl | 0.112 | 82.98 | 83.16 |
| 82 | 2,2',3,3',4-Pentachlorobiphenyl | 0.105 | 88.02 | 86.65 |
| 83 | 2,2',3,3',5-Pentachlorobiphenyl | 0.104 | 88.02 | 86.52 |
| 84 | 2,2',3,3',6-Pentachlorobiphenyl | 0.103 | 88.02 | 86.39 |
| 85 | 2,2',3,4,4'-Pentachlorobiphenyl | 0.105 | 88.02 | 86.65 |
| 86 | 2,2',3,4,5-Pentachlorobiphenyl | 0.109 | 88.02 | 87.16 |
| 87 | 2,2',3,4,5'-Pentachlorobiphenyl | 0.104 | 88.02 | 86.52 |
| 88 | 2,2',3,4,6-Pentachlorobiphenyl | 0.109 | 88.02 | 87.16 |
| 89 | 2,2',3,4,6'-Pentachlorobiphenyl | 0.102 | 88.02 | 86.26 |
| 90 | 2,2',3,4',5-Pentachlorobiphenyl | 0.105 | 88.02 | 86.65 |
| 91 | 2,2',3,4',6-Pentachlorobiphenyl | 0.104 | 88.02 | 86.52 |
| 92 | 2,2',3,5,5'-Pentachlorobiphenyl | 0.104 | 88.02 | 86.52 |
| 93 | 2,2',3,5,6-Pentachlorobiphenyl | 0.108 | 88.02 | 87.03 |
| 94 | 2,2',3,5,6'-Pentachlorobiphenyl | 0.101 | 88.02 | 86.13 |
| 96 | 2,2',3,6,6'-Pentachlorobiphenyl | 0.100 | 88.02 | 86.01 |
| 97 | 2,2',3,4',5'-Pentachlorobiphenyl | 0.106 | 88.02 | 86.77 |
| 99 | 2,2',4,4',5-Pentachlorobiphenyl | 0.107 | 88.02 | 86.90 |
| 100 | 2,2',4,4',6-Pentachlorobiphenyl | 0.106 | 88.02 | 86.77 |
| 102 | 2,2',4,5,6'-Pentachlorobiphenyl | 0.103 | 88.02 | 86.39 |
| 103 | 2,2',4,5',6-Pentachlorobiphenyl | 0.105 | 88.02 | 86.65 |
| 104 | 2,2',4,6,6'-Pentachlorobiphenyl | 0.103 | 88.02 | 86.39 |
| 105 | 2,3,3',4,4'-Pentachlorobiphenyl | 0.112 | 88.02 | 87.54 |
| 106 | 2,3,3',4,5-Pentachlorobiphenyl | 0.114 | 88.02 | 87.80 |
| 107 | 2,3,3',4',5-Pentachlorobiphenyl | 0.112 | 88.02 | 87.54 |
| 108 | 2,3,3',4,5'-Pentachlorobiphenyl | 0.108 | 88.02 | 87.03 |
| 109 | 2,3,3',4,6-Pentachlorobiphenyl | 0.112 | 88.02 | 87.54 |
| 111 | 2,3,3',5,5'-Pentachlorobiphenyl | 0.108 | 88.02 | 87.03 |
| 112 | 2,3,3',5,6-Pentachlorobiphenyl | 0.112 | 88.02 | 87.54 |
| 113 | 2,3,3',5',6-Pentachlorobiphenyl | 0.107 | 88.02 | 86.90 |
| 114 | 2,3,4,4',5-Pentachlorobiphenyl | 0.112 | 88.02 | 87.54 |
| 115 | 2,3,4,4',6-Pentachlorobiphenyl | 0.112 | 88.02 | 87.54 |
| 116 | 2,3,4,5,6-Pentachlorobiphenyl | 0.114 | 88.02 | 87.80 |
| 117 | 2,3,4',5,6-Pentachlorobiphenyl | 0.112 | 88.02 | 87.54 |
| 119 | 2,3',4,4',6-Pentachlorobiphenyl | 0.108 | 88.02 | 87.03 |
| 120 | 2,3',4,5,5'-Pentachlorobiphenyl | 0.110 | 88.02 | 87.29 |
| 121 | 2,3',4,5',6-Pentachlorobiphenyl | 0.109 | 88.02 | 87.16 |
| 122 | 2,3,3',4',5-Pentachlorobiphenyl | 0.106 | 88.02 | 86.77 |
| 123 | 2,3',4,4',5-Pentachlorobiphenyl | 0.106 | 88.02 | 86.77 |
| 124 | 2,3',4',5,5'-Pentachlorobiphenyl | 0.105 | 88.02 | 86.65 |
| 125 | 2,3',4',5',6-Pentachlorobiphenyl | 0.103 | 88.02 | 86.39 |
| 127 | 3,3',4,5,5'-Pentachlorobiphenyl | 0.109 | 88.02 | 87.16 |
| 128 | 2,2',3,3',4,4'-Hexachlorobiphenyl | 0.109 | 93.06 | 91.54 |
| 129 | 2,2',3,3',4,5-Hexachlorobiphenyl | 0.112 | 93.06 | 91.92 |
| 130 | 2,2',3,3',4,5'-Hexachlorobiphenyl | 0.108 | 93.06 | 91.41 |
| 131 | 2,2',3,3',4,6-Hexachlorobiphenyl | 0.112 | 93.06 | 91.92 |
| 132 | 2,2',3,3',4,6'-Hexachlorobiphenyl | 0.107 | 93.06 | 91.28 |
| 133 | 2,2',3,3',5,5'-Hexachlorobiphenyl | 0.108 | 93.06 | 91.41 |
| 134 | 2,2',3,3',5,6-Hexachlorobiphenyl | 0.111 | 93.06 | 91.80 |
| 135 | 2,2',3,3',5,6'-Hexachlorobiphenyl | 0.106 | 93.06 | 91.16 |
| 136 | 2,2',3,3',6,6'-Hexachlorobiphenyl | 0.103 | 93.06 | 90.77 |
| 138 | 2,2',3,4,4',5-Hexachlorobiphenyl | 0.109 | 93.06 | 91.54 |
| 139 | 2,2',3,4,4',6-Hexachlorobiphenyl | 0.113 | 93.06 | 92.05 |
| 140 | 2,2',3,4,4',6'-Hexachlorobiphenyl | 0.109 | 93.06 | 91.54 |
| 141 | 2,2',3,4,5,5'-Hexachlorobiphenyl | 0.112 | 93.06 | 91.92 |
| 142 | 2,2',3,4,5,6-Hexachlorobiphenyl | 0.114 | 93.06 | 92.18 |
| 143 | 2,2',3,4,5,6'-Hexachlorobiphenyl | 0.109 | 93.06 | 91.54 |
| 144 | 2,2',3,4,5',6-Hexachlorobiphenyl | 0.112 | 93.06 | 91.92 |
| 145 | 2,2',3,4,6,6'-Hexachlorobiphenyl | 0.109 | 93.06 | 91.54 |
| 146 | 2,2',3,4',5,5'-Hexachlorobiphenyl | 0.110 | 93.06 | 91.67 |
| 147 | 2,2',3,4',5,6-Hexachlorobiphenyl | 0.112 | 93.06 | 91.92 |
| 148 | 2,2',3,4',5,6'-Hexachlorobiphenyl | 0.108 | 93.06 | 91.41 |
| 150 | 2,2',3,4',6,6'-Hexachlorobiphenyl | 0.106 | 93.06 | 91.16 |
| 151 | 2,2',3,5,5',6-Hexachlorobiphenyl | 0.111 | 93.06 | 91.80 |
| 152 | 2,2',3,5,6,6'-Hexachlorobiphenyl | 0.108 | 93.06 | 91.41 |
| 154 | 2,2',4,4',5,6'-Hexachlorobiphenyl | 0.108 | 93.06 | 91.41 |
| 155 | 2,2',4,4',6,6'-Hexachlorobiphenyl | 0.108 | 93.06 | 91.41 |
| 156 | 2,3,3',4,4',5-Hexachlorobiphenyl | 0.119 | 93.06 | 92.82 |
| 157 | 2,3,3',4,4',5'-Hexachlorobiphenyl | 0.111 | 93.06 | 91.80 |
| 158 | 2,3,3',4,4',6-Hexachlorobiphenyl | 0.115 | 93.06 | 92.31 |
| 159 | 2,3,3',4,5,5'-Hexachlorobiphenyl | 0.116 | 93.06 | 92.44 |
| 160 | 2,3,3',4,5,6-Hexachlorobiphenyl | 0.118 | 93.06 | 92.69 |
| 161 | 2,3,3',4,5',6-Hexachlorobiphenyl | 0.116 | 93.06 | 92.44 |
| 162 | 2,3,3',4',5,5'-Hexachlorobiphenyl | 0.111 | 93.06 | 91.80 |
| 163 | 2,3,3',4',5,6-Hexachlorobiphenyl | 0.115 | 93.06 | 92.31 |
| 164 | 2,3,3',4',5',6-Hexachlorobiphenyl | 0.110 | 93.06 | 91.67 |
| 165 | 2,3,3',5,5',6-Hexachlorobiphenyl | 0.115 | 93.06 | 92.31 |
| 166 | 2,3,4,4',5,6-Hexachlorobiphenyl | 0.118 | 93.06 | 92.69 |
| 167 | 2,3',4,4',5,5'-Hexachlorobiphenyl | 0.112 | 93.06 | 91.92 |
| 168 | 2,3',4,4',5',6-Hexachlorobiphenyl | 0.112 | 93.06 | 91.92 |
| 169 | 3,3',4,4',5,5'-Hexachlorobiphenyl | 0.112 | 93.06 | 91.92 |
| 171 | 2,2',3,3',4,4',6-Heptachlorobiphenyl | 0.116 | 98.10 | 96.82 |
| 172 | 2,2',3,3',4,5,5'-Heptachlorobiphenyl | 0.116 | 98.10 | 96.82 |
| 173 | 2,2',3,3',4,5,6-Heptachlorobiphenyl | 0.118 | 98.10 | 97.08 |
| 175 | 2,2',3,3',4,5',6-Heptachlorobiphenyl | 0.115 | 98.10 | 96.69 |
| 176 | 2,2',3,3',4,6,6'-Heptachlorobiphenyl | 0.112 | 98.10 | 96.31 |
| 177 | 2,2',3,3',4,5',6'-Heptachlorobiphenyl | 0.114 | 98.10 | 96.56 |
| 178 | 2,2',3,3',5,5',6-Heptachlorobiphenyl | 0.114 | 98.10 | 96.56 |
| 179 | 2,2',3,3',5,6,6'-Heptachlorobiphenyl | 0.111 | 98.10 | 96.18 |
| 181 | 2,2',3,4,4',5,6-Heptachlorobiphenyl | 0.119 | 98.10 | 97.20 |
| 182 | 2,2',3,4,4',5,6'-Heptachlorobiphenyl | 0.113 | 98.10 | 96.44 |
| 183 | 2,2',3,4,4',5',6-Heptachlorobiphenyl | 0.115 | 98.10 | 96.69 |
| 184 | 2,2',3,4,4',6,6'-Heptachlorobiphenyl | 0.113 | 98.10 | 96.44 |
| 185 | 2,2',3,4,5,5',6-Heptachlorobiphenyl | 0.118 | 98.10 | 97.08 |
| 186 | 2,2',3,4,5,6,6'-Heptachlorobiphenyl | 0.115 | 98.10 | 96.69 |
| 187 | 2,2',3,4',5,5',6-Heptachlorobiphenyl | 0.114 | 98.10 | 96.56 |
| 188 | 2,2',3,4',5,6,6'-Heptachlorobiphenyl | 0.111 | 98.10 | 96.18 |
| 189 | 2,3,3',4,4',5,5'-Heptachlorobiphenyl | 0.119 | 98.10 | 97.20 |
| 190 | 2,3,3',4,4',5,6-Heptachlorobiphenyl | 0.121 | 98.10 | 97.46 |
| 191 | 2,3,3',4,4',5',6-Heptachlorobiphenyl | 0.119 | 98.10 | 97.20 |
| 192 | 2,3,3',4,5,5',6-Heptachlorobiphenyl | 0.122 | 98.10 | 97.59 |
| 193 | 2,3,3',4',5,5',6-Heptachlorobiphenyl | 0.118 | 98.10 | 97.08 |
| 194 | 2,2',3,3',4,4',5,5'-Octachlorobiphenyl | 0.119 | 103.10 | 101.60 |
| 195 | 2,2',3,3',4,4',5,6-Octachlorobiphenyl | 0.121 | 103.10 | 101.80 |
| 196 | 2,2',3,3',4,4',5,6'-Octachlorobiphenyl | 0.118 | 103.10 | 101.50 |
| 197 | 2,2',3,3',4,4',6,6'-Octachlorobiphenyl | 0.116 | 103.10 | 101.20 |
| 198 | 2,2',3,3',4,5,5',6-Octachlorobiphenyl | 0.121 | 103.10 | 101.80 |
| 199 | 2,2',3,3',4,5,5',6'-Octachlorobiphenyl | 0.117 | 103.10 | 101.30 |
| 200 | 2,2',3,3',4,5,6,6'-Octachlorobiphenyl | 0.118 | 103.10 | 101.50 |
| 201 | 2,2',3,3',4,5',6,6'-Octachlorobiphenyl | 0.115 | 103.10 | 101.10 |
| 202 | 2,2',3,3',5,5',6,6'-Octachlorobiphenyl | 0.114 | 103.10 | 100.90 |
| 203 | 2,2',3,4,4',5,5',6-Octachlorobiphenyl | 0.121 | 103.10 | 101.80 |
| 204 | 2,2',3,4,4',5,6,6'-Octachlorobiphenyl | 0.119 | 103.10 | 101.60 |
| 205 | 2,3,3',4,4',5,5',6-Octachlorobiphenyl | 0.125 | 103.10 | 102.40 |
| 206 | 2,2',3,3',4,4',5,5',6-Nonachlorobiphenyl | 0.124 | 108.20 | 106.60 |
| 207 | 2,2',3,3',4,4',5,6,6'-Nonachlorobiphenyl | 0.122 | 108.20 | 106.40 |
| 208 | 2,2',3,3',4,5,5',6,6'-Nonachlorobiphenyl | 0.121 | 108.20 | 106.20 |
| 209 | 2,2',3,3',4,4',5,5',6,6'-Decachlorobiphenyl | 0.124 | 108.20 | 106.60 |

[a]Calculated $\Delta_{vap}H_m$ using one parameter ($N_{NH}$) regression.
[b]Calculated $\Delta_{vap}H_m$ using two parameter ($\omega$, $N_{NH}$) regression.

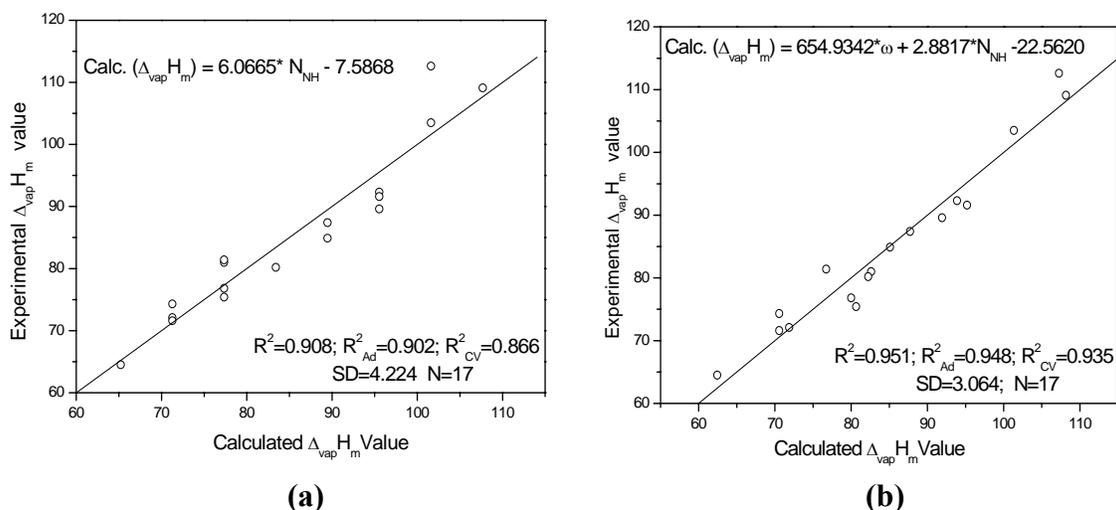

**Figure 3**: Calculated vs experimental values of $\Delta_{vap}H_m$ of 17 PCBs using (a) one parameter ($N_{NH}$) and (b) two parameter ($N_{NH}$, $\omega$) regression.

Regression Equations:

| Parameters | Regression Equations | $R^2$ | $R^2_{(CV)}$ | $R^2_{(Ad)}$ |
|---|---|---|---|---|
| One | Calc. ($\Delta_{vap}H_m$)=6.0665* $N_{NH}$ - 7.5868 | 0.908 | 0.866 | 0.902 |
| Two | Calc. ($\Delta_{vap}H_m$) =654.9342*$\omega$ +2.8817* $N_{NH}$ -22.5620 | 0.951 | 0.935 | 0.948 |

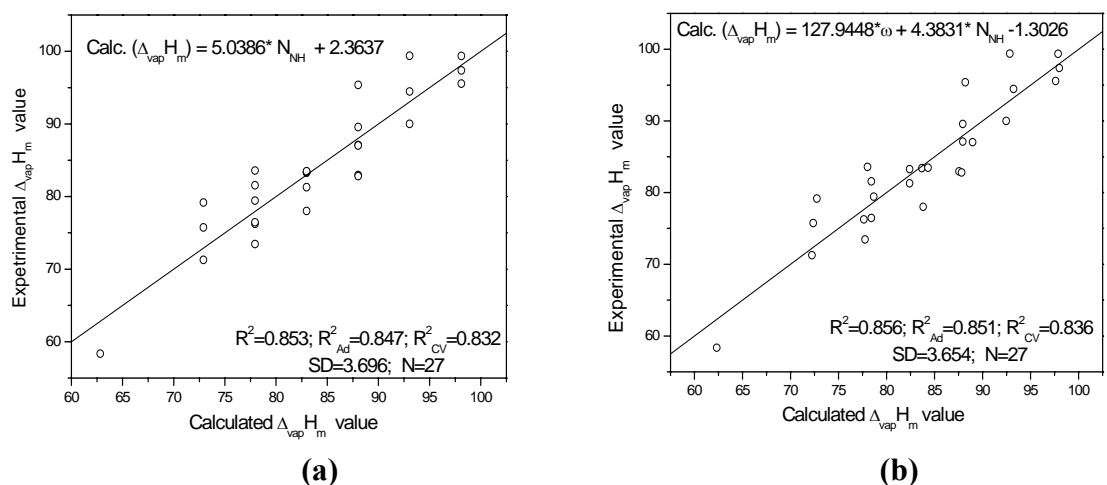

**Figure 4:** Calculated vs experimental values of $\Delta_{vap}H_m$ of 27 PCBs using a) one parameter ($N_{NH}$) and b) two parameter ($N_{NH}$, $\omega$) regressions.

Regression Equations:

| Parameters | Regression Equations | $R^2$ | $R^2_{(CV)}$ | $R^2_{(Ad)}$ |
|---|---|---|---|---|
| One | Calc. ($\Delta_{vap}H_m$) = 5.0386* $N_{NH}$ + 2.3637 | 0.853 | 0.832 | 0.847 |
| Two | Calc. ($\Delta_{vap}H_m$) = 127.9448* $\omega$ + 4.3831* $N_{NH}$ - 1.3026 | 0.856 | 0.836 | 0.851 |

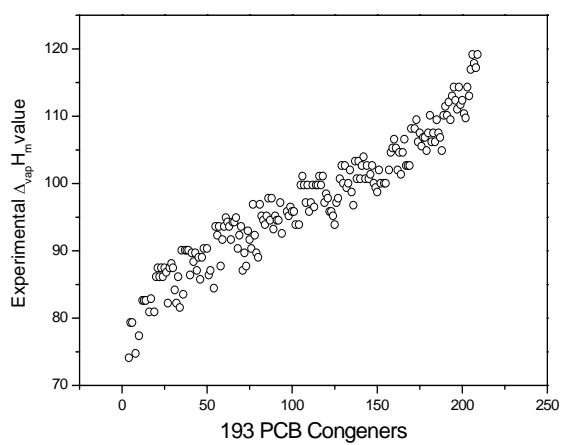 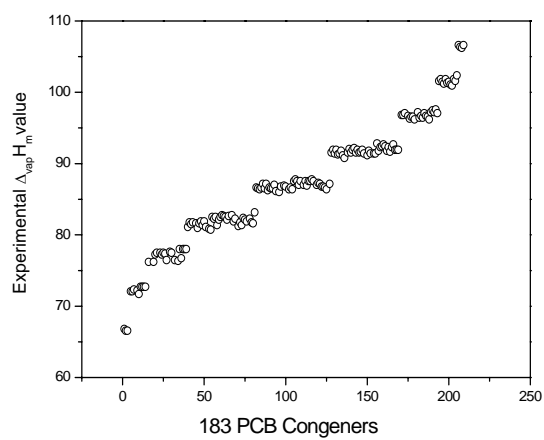

(a) **<u>Using 17 PCB</u>**   (b) **<u>Using 27 PCB</u>**

**Figure 5:** Predicted $\Delta_{vap}H_m$ values in the general set for (a) 193 PCB congeners and (b) 183 PCB congeners.

**Table 6.** QSPR models for log P calculation of 100 PCBs as training set.

| IUPAC No | Compounds | $\omega$ | log P Value | | |
|---|---|---|---|---|---|
| | | | Exptl* | Calcd[a] ($N_{NH}$) | Calcd[b] ($\omega$, $N_{NH}$) |
| 2 | 3-Chlorobiphenyl | 0.085 | 4.66 | 4.73 | 4.52 |
| 11 | 3,3'-Dichlorobiphenyl | 0.099 | 5.27 | 5.13 | 5.37 |
| 12 | 3,4-Dichlorobiphenyl | 0.099 | 5.23 | 5.13 | 5.37 |
| 13 | 3,4'-Dichlorobiphenyl | 0.099 | 5.15 | 5.13 | 5.37 |
| 15 | 4,4'-Dichlorobiphenyl | 0.090 | 5.23 | 5.13 | 4.94 |
| 16 | 2,2',3-Trichlorobiphenyl | 0.092 | 5.12 | 5.52 | 5.22 |
| 22 | 2,3,4'-Trichlorobiphenyl | 0.100 | 5.29 | 5.52 | 5.60 |
| 24 | 2,3,6-Trichlorobiphenyl | 0.100 | 5.44 | 5.52 | 5.60 |
| 25 | 2,3',4-Trichlorobiphenyl | 0.102 | 5.54 | 5.52 | 5.69 |
| 26 | 2,3',5-Trichlorobiphenyl | 0.101 | 5.65 | 5.52 | 5.65 |
| 28 | 2,4,4'-Trichlorobiphenyl | 0.102 | 5.71 | 5.52 | 5.69 |
| 31 | 2,4',5-Trichlorobiphenyl | 0.097 | 5.68 | 5.52 | 5.46 |
| 32 | 2,4',6-Trichlorobiphenyl | 0.094 | 5.24 | 5.52 | 5.32 |
| 33 | 2,3',4'-Trichlorobiphenyl | 0.100 | 5.71 | 5.52 | 5.60 |
| 34 | 2,3',5'-Trichlorobiphenyl | 0.093 | 5.71 | 5.52 | 5.27 |
| 40 | 2,2',3,3'-Tetrachlorobiphenyl | 0.096 | 5.67 | 5.92 | 5.60 |
| 41 | 2,2',3,4-Tetrachlorobiphenyl | 0.101 | 5.79 | 5.92 | 5.83 |
| 47 | 2,2',4,4'-Tetrachlorobiphenyl | 0.100 | 5.94 | 5.92 | 5.79 |
| 48 | 2,2',4,5-Tetrachlorobiphenyl | 0.102 | 5.69 | 5.92 | 5.88 |
| 49 | 2,2',4,5'-Tetrachlorobiphenyl | 0.098 | 5.87 | 5.92 | 5.69 |
| 50 | 2,2',4,6-Tetrachlorobiphenyl | 0.102 | 5.75 | 5.92 | 5.88 |
| 51 | 2,2',4,6'-Tetrachlorobiphenyl | 0.096 | 5.51 | 5.92 | 5.60 |
| 52 | 2,2',5,5'-Tetrachlorobiphenyl | 0.097 | 5.79 | 5.92 | 5.65 |
| 53 | 2,2',5,6'-Tetrachlorobiphenyl | 0.094 | 5.55 | 5.92 | 5.51 |
| 54 | 2,2',6,6'-Tetrachlorobiphenyl | 0.093 | 5.24 | 5.92 | 5.46 |
| 55 | 2,3,3',4-Tetrachlorobiphenyl | 0.107 | 6.1 | 5.92 | 6.12 |
| 60 | 2,3,4,4'-Tetrachlorobiphenyl | 0.107 | 6.24 | 5.92 | 6.12 |
| 63 | 2,3,4',5-Tetrachlorobiphenyl | 0.107 | 6.1 | 5.92 | 6.12 |
| 64 | 2,3,4',6-Tetrachlorobiphenyl | 0.104 | 5.76 | 5.92 | 5.97 |
| 65 | 2,3,5,6-Tetrachlorobiphenyl | 0.108 | 5.96 | 5.92 | 6.16 |
| 66 | 2,3',4,4'-Tetrachlorobiphenyl | 0.108 | 5.98 | 5.92 | 6.16 |
| 67 | 2,3',4,5-Tetrachlorobiphenyl | 0.109 | 6.32 | 5.92 | 6.21 |
| 69 | 2,3',4,6-Tetrachlorobiphenyl | 0.105 | 6.03 | 5.92 | 6.02 |
| 71 | 2,3',4',6-Tetrachlorobiphenyl | 0.097 | 5.76 | 5.92 | 5.65 |
| 74 | 2,4,4',5-Tetrachlorobiphenyl | 0.106 | 6.1 | 5.92 | 6.07 |
| 75 | 2,4,4',6-Tetrachlorobiphenyl | 0.104 | 6.03 | 5.92 | 5.97 |
| 76 | 2,3',4',5'-Tetrachlorobiphenyl | 0.102 | 5.98 | 5.92 | 5.88 |
| 84 | 2,2',3,3',6-Pentachlorobiphenyl | 0.103 | 5.6 | 6.31 | 6.11 |
| 90 | 2,2',3,4',5-Pentachlorobiphenyl | 0.105 | 6.32 | 6.31 | 6.21 |
| 91 | 2,2',3,4',6-Pentachlorobiphenyl | 0.104 | 5.87 | 6.31 | 6.16 |
| 92 | 2,2',3,5,5'-Pentachlorobiphenyl | 0.104 | 6.32 | 6.31 | 6.16 |
| 93 | 2,2',3,5,6-Pentachlorobiphenyl | 0.108 | 6.06 | 6.31 | 6.35 |
| 95 | 2,2',3,5',6-Pentachlorobiphenyl | 0.103 | 5.92 | 6.31 | 6.11 |
| 97 | 2,2',3,4',5'-Pentachlorobiphenyl | 0.106 | 6.3 | 6.31 | 6.25 |
| 98 | 2,2',3,4',6'-Pentachlorobiphenyl | 0.105 | 6.04 | 6.31 | 6.21 |
| 99 | 2,2',4,4',5-Pentachlorobiphenyl | 0.107 | 6.41 | 6.31 | 6.30 |
| 100 | 2,2',4,4',6-Pentachlorobiphenyl | 0.106 | 6.23 | 6.31 | 6.25 |
| 103 | 2,2',4,5',6-Pentachlorobiphenyl | 0.105 | 6.11 | 6.31 | 6.21 |
| 105 | 2,3,3',4,4'-Pentachlorobiphenyl | 0.112 | 6.79 | 6.31 | 6.54 |
| 106 | 2,3,3',4,5-Pentachlorobiphenyl | 0.114 | 6.92 | 6.31 | 6.63 |
| 110 | 2,3,3',4',6-Pentachlorobiphenyl | 0.106 | 6.2 | 6.31 | 6.25 |
| 112 | 2,3,3',5,6-Pentachlorobiphenyl | 0.112 | 6.41 | 6.31 | 6.54 |
| 113 | 2,3,3',5',6-Pentachlorobiphenyl | 0.107 | 6.45 | 6.31 | 6.30 |
| 114 | 2,3,4,4',5-Pentachlorobiphenyl | 0.112 | 6.71 | 6.31 | 6.54 |
| 115 | 2,3,4,4',6-Pentachlorobiphenyl | 0.112 | 6.44 | 6.31 | 6.54 |
| 117 | 2,3,4',5,6-Pentachlorobiphenyl | 0.112 | 6.39 | 6.31 | 6.54 |
| 118 | 2,3',4,4',5-Pentachlorobiphenyl | 0.114 | 6.57 | 6.31 | 6.63 |
| 119 | 2,3',4,4',6-Pentachlorobiphenyl | 0.108 | 6.4 | 6.31 | 6.35 |
| 120 | 2,3',4,5,5'-Pentachlorobiphenyl | 0.110 | 6.3 | 6.31 | 6.44 |
| 121 | 2,3',4,5',6-Pentachlorobiphenyl | 0.109 | 6.42 | 6.31 | 6.40 |
| 123 | 2,3',4,4',5'-Pentachlorobiphenyl | 0.106 | 6.64 | 6.31 | 6.25 |
| 135 | 2,2',3,3',5,6'-Hexachlorobiphenyl | 0.106 | 6.32 | 6.71 | 6.44 |
| 137 | 2,2',3,4,4',5-Hexachlorobiphenyl | 0.113 | 6.82 | 6.71 | 6.77 |
| 138 | 2,2',3,4,4',5'-Hexachlorobiphenyl | 0.109 | 6.73 | 6.71 | 6.58 |
| 140 | 2,2',3,4,4',6'-Hexachlorobiphenyl | 0.109 | 6.58 | 6.71 | 6.58 |
| 141 | 2,2',3,4,5,5'-Hexachlorobiphenyl | 0.112 | 6.75 | 6.71 | 6.72 |
| 143 | 2,2',3,4,5,6'-Hexachlorobiphenyl | 0.109 | 6.56 | 6.71 | 6.58 |
| 144 | 2,2',3,4,5',6-Hexachlorobiphenyl | 0.112 | 6.45 | 6.71 | 6.72 |
| 146 | 2,2',3,4',5,5'-Hexachlorobiphenyl | 0.11 | 6.85 | 6.71 | 6.63 |
| 149 | 2,2',3,4',5',6-Hexachlorobiphenyl | 0.107 | 6.41 | 6.71 | 6.49 |
| 151 | 2,2',3,5,5',6-Hexachlorobiphenyl | 0.111 | 6.42 | 6.71 | 6.68 |
| 153 | 2,2',4,4',5,5'-Hexachlorobiphenyl | 0.110 | 6.8 | 6.71 | 6.63 |
| 154 | 2,2',4,4',5,6'-Hexachlorobiphenyl | 0.108 | 6.65 | 6.71 | 6.54 |
| 155 | 2,2',4,4',6,6'-Hexachlorobiphenyl | 0.108 | 6.54 | 6.71 | 6.54 |
| 156 | 2,3,3',4,4',5-Hexachlorobiphenyl | 0.119 | 7.44 | 6.71 | 7.05 |
| 158 | 2,3,3',4,4',6-Hexachlorobiphenyl | 0.115 | 6.78 | 6.71 | 6.86 |
| 163 | 2,3,3',4',5,6-Hexachlorobiphenyl | 0.115 | 6.78 | 6.71 | 6.86 |

| 164 | 2,3,3',4',5',6-Hexachlorobiphenyl | 0.110 | 6.63 | 6.71 | 6.63 |
| 165 | 2,3,3',5,5',6-Hexachlorobiphenyl | 0.115 | 7 | 6.71 | 6.86 |
| 167 | 2,3',4,4',5,5'-Hexachlorobiphenyl | 0.112 | 7.29 | 6.71 | 6.72 |
| 169 | 3,3',4,4',5,5'-Hexachlorobiphenyl | 0.112 | 7.55 | 6.71 | 6.72 |
| 170 | 2,2',3,3',4,4',5-Heptachlorobiphenyl | 0.116 | 7.08 | 7.10 | 7.10 |
| 177 | 2,2',3,3',4,5',6'-Heptachlorobiphenyl | 0.114 | 6.73 | 7.10 | 7.00 |
| 178 | 2,2',3,3',5,5',6-Heptachlorobiphenyl | 0.114 | 6.85 | 7.10 | 7.00 |
| 179 | 2,2',3,3',5,6,6'-Heptachlorobiphenyl | 0.111 | 6.41 | 7.10 | 6.86 |
| 180 | 2,2',3,4,4',5,5'-Heptachlorobiphenyl | 0.115 | 7.21 | 7.10 | 7.05 |
| 181 | 2,2',3,4,4',5,6-Heptachlorobiphenyl | 0.119 | 7.13 | 7.10 | 7.24 |
| 182 | 2,2',3,4,4',5,6'-Heptachlorobiphenyl | 0.113 | 6.92 | 7.10 | 6.96 |
| 183 | 2,2',3,4,4',5',6-Heptachlorobiphenyl | 0.115 | 7.04 | 7.10 | 7.05 |
| 185 | 2,2',3,4,5,5',6-Heptachlorobiphenyl | 0.118 | 6.99 | 7.10 | 7.19 |
| 188 | 2,2',3,4',5,6,6'-Heptachlorobiphenyl | 0.111 | 6.78 | 7.10 | 6.86 |
| 189 | 2,3,3',4,4',5,5'-Heptachlorobiphenyl | 0.119 | 7.72 | 7.10 | 7.24 |
| 190 | 2,3,3',4,4',5,6-Heptachlorobiphenyl | 0.121 | 7.08 | 7.10 | 7.33 |
| 191 | 2,3,3',4,4',5',6-Heptachlorobiphenyl | 0.119 | 7.21 | 7.10 | 7.24 |
| 192 | 2,3,3',4,5,5',6-Heptachlorobiphenyl | 0.122 | 7.21 | 7.10 | 7.38 |
| 193 | 2,3,3',4',5,5',6-Heptachlorobiphenyl | 0.118 | 7.21 | 7.10 | 7.19 |
| 194 | 2,2',3,3',4,4',5,5'-Octachlorobiphenyl | 0.119 | 7.62 | 7.50 | 7.42 |
| 195 | 2,2',3,3',4,4',5,6-Octachlorobiphenyl | 0.121 | 7.35 | 7.50 | 7.52 |
| 205 | 2,3,3',4,4',5,5',6-Octachlorobiphenyl | 0.125 | 7.62 | 7.50 | 7.71 |
| 207 | 2,2',3,3',4,4',5,6,6'-Nonachlorobiphenyl | 0.122 | 7.88 | 7.89 | 7.75 |

*Experimental data taken from References [19,20].
[a]Calculated log P values using one parameter ($N_{NH}$) regression.
[b]Calculated log P values using two parameter ($\omega$, $N_{NH}$) regression.

**Table 7:** Calculation of log P values of 33 PCBs as test set

| IUPAC No | Compounds | $\omega$ | log P Value | | |
| --- | --- | --- | --- | --- | --- |
| | | | Expl* | Calcd[a] ($N_{NH}$) | Calcd[b] ($\omega$, $N_{NH}$) |
| 3 | 4-Chlorobiphenyl | 0.085 | 4.63 | 4.73 | 4.52 |
| 4 | 2,2'-Dichlorobiphenyl | 0.086 | 4.72 | 5.13 | 4.76 |
| 5 | 2,3-Dichlorobiphenyl | 0.094 | 4.99 | 5.13 | 5.13 |
| 6 | 2,3'-Dichlorobiphenyl | 0.094 | 4.84 | 5.13 | 5.13 |
| 7 | 2,4-Dichlorobiphenyl | 0.096 | 5.15 | 5.13 | 5.23 |
| 8 | 2,4'-Dichlorobiphenyl | 0.087 | 5.09 | 5.13 | 4.80 |
| 17 | 2,2',4-Trichlorobiphenyl | 0.095 | 5.39 | 5.52 | 5.37 |
| 18 | 2,20,5-Trichlorobiphenyl | 0.094 | 5.33 | 5.52 | 5.32 |
| 19 | 2,2',6-Trichlorobiphenyl | 0.092 | 5.04 | 5.52 | 5.23 |
| 20 | 2,3,3'-Trichlorobiphenyl | 0.100 | 5.6 | 5.52 | 5.60 |
| 21 | 2,3,4-Trichlorobiphenyl | 0.102 | 5.68 | 5.52 | 5.70 |
| 42 | 2,2',3,4'-Tetrachlorobiphenyl | 0.099 | 5.72 | 5.92 | 5.74 |
| 44 | 2,2',3,5'-Tetrachlorobiphenyl | 0.097 | 5.73 | 5.92 | 5.65 |
| 45 | 2,2',3,6-Tetrachlorobiphenyl | 0.100 | 4.84 | 5.92 | 5.79 |
| 46 | 2,2',3,6'-Tetrachlorobiphenyl | 0.095 | 4.84 | 5.92 | 5.55 |
| 85 | 2,2',3,4,4'-Pentachlorobiphenyl | 0.105 | 6.18 | 6.31 | 6.21 |
| 86 | 2,2',3,4,5-Pentachlorobiphenyl | 0.109 | 6.38 | 6.31 | 6.40 |
| 87 | 2,2',3,4,5'-Pentachlorobiphenyl | 0.104 | 6.23 | 6.31 | 6.16 |
| 88 | 2,2',3,4,6-Pentachlorobiphenyl | 0.109 | 6.5 | 6.31 | 6.40 |

| 89 | 2,2',3,4,6'-Pentachlorobiphenyl | 0.102 | 5.6 | 6.31 | 6.07 |
| 129 | 2,2',3,3',4,5-Hexachlorobiphenyl | 0.112 | 6.76 | 6.71 | 6.72 |
| 130 | 2,2',3,3',4,5'-Hexachlorobiphenyl | 0.108 | 7.3 | 6.71 | 6.54 |
| 131 | 2,2',3,3',4,6-Hexachlorobiphenyl | 0.112 | 6.78 | 6.71 | 6.72 |
| 132 | 2,2',3,3',4,6'-Hexachlorobiphenyl | 0.107 | 6.2 | 6.71 | 6.49 |
| 134 | 2,2',3,3',5,6-Hexachlorobiphenyl | 0.111 | 6.2 | 6.71 | 6.68 |
| 172 | 2,2',3,3',4,5,5'-Heptachlorobiphenyl | 0.116 | 7.21 | 7.10 | 7.10 |
| 174 | 2,2',3,3',4,5,6'-Heptachlorobiphenyl | 0.113 | 6.85 | 7.10 | 6.96 |
| 175 | 2,2',3,3',4,5',6-Heptachlorobiphenyl | 0.115 | 6.92 | 7.10 | 7.05 |
| 176 | 2,2',3,3',4,6,6'-Heptachlorobiphenyl | 0.112 | 6.55 | 7.10 | 6.91 |
| 196 | 2,2',3,3',4,4',5,6'-Octachlorobiphenyl | 0.118 | 7.43 | 7.50 | 7.38 |
| 203 | 2,2',3,4,4',5,5',6-Octachlorobiphenyl | 0.121 | 7.49 | 7.50 | 7.52 |
| 204 | 2,2',3,4,4',5,6,6'-Octachlorobiphenyl | 0.119 | 7.48 | 7.50 | 7.43 |
| 206 | 2,2',3,3',4,4',5,5',6-Nonachlorobiphenyl | 0.124 | 7.94 | 7.89 | 7.85 |

*Experimental data taken from References [19,20].
[a] Calculated log P values using one parameter ($N_{NH}$) regression.
[b] Calculated log P values using two parameter ($\omega$, $N_{NH}$) regression.

**Table 8:** Calculation of log P values of 133 PCBs as the data set:

| IUPAC No | Compounds | $\omega$ | log P Value | | |
|---|---|---|---|---|---|
| | | | Exptl* | Calcd[a] ($N_{NH}$) | Calcd[b] ($\omega$, $N_{NH}$) |
| 2 | 3-Chlorobiphenyl | 0.085 | 4.66 | 4.64 | 4.47 |
| 3 | 4-Chlorobiphenyl | 0.085 | 4.63 | 4.64 | 4.47 |
| 4 | 2,2'-Dichlorobiphenyl | 0.086 | 4.72 | 5.05 | 4.69 |
| 5 | 2,3-Dichlorobiphenyl | 0.094 | 4.99 | 5.05 | 5.10 |
| 6 | 2,3'-Dichlorobiphenyl | 0.094 | 4.84 | 5.05 | 5.10 |
| 7 | 2,4-Dichlorobiphenyl | 0.096 | 5.15 | 5.05 | 5.21 |
| 8 | 2,4'-Dichlorobiphenyl | 0.087 | 5.09 | 5.05 | 4.75 |
| 11 | 3,3'-Dichlorobiphenyl | 0.099 | 5.27 | 5.05 | 5.36 |
| 12 | 3,4-Dichlorobiphenyl | 0.099 | 5.23 | 5.05 | 5.36 |
| 13 | 3,4'-Dichlorobiphenyl | 0.099 | 5.15 | 5.05 | 5.36 |
| 15 | 4,4'-Dichlorobiphenyl | 0.09 | 5.23 | 5.05 | 4.90 |
| 16 | 2,2',3-Trichlorobiphenyl | 0.092 | 5.12 | 5.45 | 5.17 |
| 17 | 2,2',4-Trichlorobiphenyl | 0.095 | 5.39 | 5.45 | 5.33 |
| 18 | 2,20,5-Trichlorobiphenyl | 0.094 | 5.33 | 5.45 | 5.28 |
| 19 | 2,2',6-Trichlorobiphenyl | 0.092 | 5.04 | 5.45 | 5.17 |
| 20 | 2,3,3'-Trichlorobiphenyl | 0.100 | 5.6 | 5.45 | 5.58 |
| 21 | 2,3,4-Trichlorobiphenyl | 0.102 | 5.68 | 5.45 | 5.68 |
| 22 | 2,3,4'-Trichlorobiphenyl | 0.100 | 5.29 | 5.45 | 5.58 |
| 24 | 2,3,6-Trichlorobiphenyl | 0.100 | 5.44 | 5.45 | 5.58 |
| 25 | 2,3',4-Trichlorobiphenyl | 0.102 | 5.54 | 5.45 | 5.68 |
| 26 | 2,3',5-Trichlorobiphenyl | 0.101 | 5.65 | 5.45 | 5.63 |
| 28 | 2,4,4'-Trichlorobiphenyl | 0.102 | 5.71 | 5.45 | 5.68 |
| 31 | 2,4',5-Trichlorobiphenyl | 0.097 | 5.68 | 5.45 | 5.43 |
| 32 | 2,4',6-Trichlorobiphenyl | 0.094 | 5.24 | 5.45 | 5.28 |
| 33 | 2,3',4'-Trichlorobiphenyl | 0.100 | 5.71 | 5.45 | 5.58 |
| 34 | 2,3',5'-Trichlorobiphenyl | 0.093 | 5.71 | 5.45 | 5.22 |
| 40 | 2,2',3,3'-Tetrachlorobiphenyl | 0.096 | 5.67 | 5.86 | 5.55 |
| 41 | 2,2',3,4-Tetrachlorobiphenyl | 0.101 | 5.79 | 5.86 | 5.81 |
| 42 | 2,2',3,4'-Tetrachlorobiphenyl | 0.099 | 5.72 | 5.86 | 5.70 |
| 44 | 2,2',3,5'-Tetrachlorobiphenyl | 0.097 | 5.73 | 5.86 | 5.60 |
| 45 | 2,2',3,6-Tetrachlorobiphenyl | 0.100 | 4.84 | 5.86 | 5.75 |
| 46 | 2,2',3,6'-Tetrachlorobiphenyl | 0.095 | 4.84 | 5.86 | 5.50 |
| 47 | 2,2',4,4'-Tetrachlorobiphenyl | 0.100 | 5.94 | 5.86 | 5.75 |
| 48 | 2,2',4,5-Tetrachlorobiphenyl | 0.102 | 5.69 | 5.86 | 5.86 |
| 49 | 2,2',4,5'-Tetrachlorobiphenyl | 0.098 | 5.87 | 5.86 | 5.65 |
| 50 | 2,2',4,6-Tetrachlorobiphenyl | 0.102 | 5.75 | 5.86 | 5.86 |
| 51 | 2,2',4,6'-Tetrachlorobiphenyl | 0.096 | 5.51 | 5.86 | 5.55 |
| 52 | 2,2',5,5'-Tetrachlorobiphenyl | 0.097 | 5.79 | 5.86 | 5.60 |
| 53 | 2,2',5,6'-Tetrachlorobiphenyl | 0.094 | 5.55 | 5.86 | 5.45 |
| 54 | 2,2',6,6'-Tetrachlorobiphenyl | 0.093 | 5.24 | 5.86 | 5.40 |
| 55 | 2,3,3',4-Tetrachlorobiphenyl | 0.107 | 6.1 | 5.86 | 6.11 |
| 60 | 2,3,4,4'-Tetrachlorobiphenyl | 0.107 | 6.24 | 5.86 | 6.11 |
| 63 | 2,3,4',5-Tetrachlorobiphenyl | 0.107 | 6.1 | 5.86 | 6.11 |
| 64 | 2,3,4',6-Tetrachlorobiphenyl | 0.104 | 5.76 | 5.86 | 5.96 |
| 65 | 2,3,5,6-Tetrachlorobiphenyl | 0.108 | 5.96 | 5.86 | 6.16 |
| 66 | 2,3',4,4'-Tetrachlorobiphenyl | 0.108 | 5.98 | 5.86 | 6.16 |
| 67 | 2,3',4,5-Tetrachlorobiphenyl | 0.109 | 6.32 | 5.86 | 6.21 |
| 69 | 2,3',4,6-Tetrachlorobiphenyl | 0.105 | 6.03 | 5.86 | 6.01 |
| 71 | 2,3',4',6-Tetrachlorobiphenyl | 0.097 | 5.76 | 5.86 | 5.60 |
| 74 | 2,4,4',5-Tetrachlorobiphenyl | 0.106 | 6.1 | 5.86 | 6.06 |
| 75 | 2,4,4',6-Tetrachlorobiphenyl | 0.104 | 6.03 | 5.86 | 5.96 |
| 76 | 2,3',4',5'-Tetrachlorobiphenyl | 0.102 | 5.98 | 5.86 | 5.86 |
| 84 | 2,2',3,3',6-Pentachlorobiphenyl | 0.103 | 5.6 | 6.27 | 6.08 |
| 85 | 2,2',3,4,4'-Pentachlorobiphenyl | 0.105 | 6.18 | 6.27 | 6.18 |
| 86 | 2,2',3,4,5-Pentachlorobiphenyl | 0.109 | 6.38 | 6.27 | 6.39 |
| 87 | 2,2',3,4,5'-Pentachlorobiphenyl | 0.104 | 6.23 | 6.27 | 6.13 |
| 88 | 2,2',3,4,6-Pentachlorobiphenyl | 0.109 | 6.5 | 6.27 | 6.39 |
| 89 | 2,2',3,4,6'-Pentachlorobiphenyl | 0.102 | 5.6 | 6.27 | 6.03 |
| 90 | 2,2',3,4',5-Pentachlorobiphenyl | 0.105 | 6.32 | 6.27 | 6.18 |
| 91 | 2,2',3,4',6-Pentachlorobiphenyl | 0.104 | 5.87 | 6.27 | 6.13 |
| 92 | 2,2',3,5,5'-Pentachlorobiphenyl | 0.104 | 6.32 | 6.27 | 6.13 |
| 93 | 2,2',3,5,6-Pentachlorobiphenyl | 0.108 | 6.06 | 6.27 | 6.34 |
| 95 | 2,2',3,5',6-Pentachlorobiphenyl | 0.103 | 5.92 | 6.27 | 6.08 |
| 97 | 2,2',3,4',5'-Pentachlorobiphenyl | 0.106 | 6.3 | 6.27 | 6.23 |
| 98 | 2,2',3,4',6'-Pentachlorobiphenyl | 0.105 | 6.04 | 6.27 | 6.18 |
| 99 | 2,2',4,4',5-Pentachlorobiphenyl | 0.107 | 6.41 | 6.27 | 6.28 |
| 100 | 2,2',4,4',6-Pentachlorobiphenyl | 0.106 | 6.23 | 6.27 | 6.23 |
| 103 | 2,2',4,5',6-Pentachlorobiphenyl | 0.105 | 6.11 | 6.27 | 6.18 |
| 105 | 2,3,3',4,4'-Pentachlorobiphenyl | 0.112 | 6.79 | 6.27 | 6.54 |
| 106 | 2,3,3',4,5-Pentachlorobiphenyl | 0.114 | 6.92 | 6.27 | 6.64 |
| 110 | 2,3,3',4',6-Pentachlorobiphenyl | 0.106 | 6.2 | 6.27 | 6.23 |
| 112 | 2,3,3',5,6-Pentachlorobiphenyl | 0.112 | 6.41 | 6.27 | 6.54 |
| 113 | 2,3,3',5',6-Pentachlorobiphenyl | 0.107 | 6.45 | 6.27 | 6.28 |
| 114 | 2,3,4,4',5-Pentachlorobiphenyl | 0.112 | 6.71 | 6.27 | 6.54 |
| 115 | 2,3,4,4',6-Pentachlorobiphenyl | 0.112 | 6.44 | 6.27 | 6.54 |
| 117 | 2,3,4',5,6-Pentachlorobiphenyl | 0.112 | 6.39 | 6.27 | 6.54 |
| 118 | 2,3',4,4',5-Pentachlorobiphenyl | 0.114 | 6.57 | 6.27 | 6.64 |
| 119 | 2,3',4,4',6-Pentachlorobiphenyl | 0.108 | 6.4 | 6.27 | 6.34 |
| 120 | 2,3',4,5,5'-Pentachlorobiphenyl | 0.110 | 6.3 | 6.27 | 6.44 |
| 121 | 2,3',4,5',6-Pentachlorobiphenyl | 0.109 | 6.42 | 6.27 | 6.39 |
| 123 | 2,3',4,4',5'-Pentachlorobiphenyl | 0.106 | 6.64 | 6.27 | 6.23 |
| 129 | 2,2',3,3',4,5-Hexachlorobiphenyl | 0.112 | 6.76 | 6.67 | 6.71 |
| 130 | 2,2',3,3',4,5'-Hexachlorobiphenyl | 0.108 | 7.3 | 6.67 | 6.51 |
| 131 | 2,2',3,3',4,6-Hexachlorobiphenyl | 0.112 | 6.78 | 6.67 | 6.71 |
| 132 | 2,2',3,3',4,6'-Hexachlorobiphenyl | 0.107 | 6.2 | 6.67 | 6.46 |
| 134 | 2,2',3,3',5,6-Hexachlorobiphenyl | 0.111 | 6.2 | 6.67 | 6.66 |
| 135 | 2,2',3,3',5,6'-Hexachlorobiphenyl | 0.106 | 6.32 | 6.67 | 6.41 |
| 137 | 2,2',3,4,4',5-Hexachlorobiphenyl | 0.113 | 6.82 | 6.67 | 6.76 |
| 138 | 2,2',3,4,4',5'-Hexachlorobiphenyl | 0.109 | 6.73 | 6.67 | 6.56 |
| 140 | 2,2',3,4,4',6'-Hexachlorobiphenyl | 0.109 | 6.58 | 6.67 | 6.56 |
| 141 | 2,2',3,4,5,5'-Hexachlorobiphenyl | 0.112 | 6.75 | 6.67 | 6.71 |
| 143 | 2,2',3,4,5,6'-Hexachlorobiphenyl | 0.109 | 6.56 | 6.67 | 6.56 |
| 144 | 2,2',3,4,5',6-Hexachlorobiphenyl | 0.112 | 6.45 | 6.67 | 6.71 |
| 146 | 2,2',3,4',5,5'-Hexachlorobiphenyl | 0.110 | 6.85 | 6.67 | 6.61 |
| 149 | 2,2',3,4',5',6-Hexachlorobiphenyl | 0.107 | 6.41 | 6.67 | 6.46 |
| 151 | 2,2',3,5,5',6-Hexachlorobiphenyl | 0.111 | 6.42 | 6.67 | 6.66 |
| 153 | 2,2',4,4',5,5'-Hexachlorobiphenyl | 0.110 | 6.8 | 6.67 | 6.61 |
| 154 | 2,2',4,4',5,6'-Hexachlorobiphenyl | 0.108 | 6.65 | 6.67 | 6.51 |
| 155 | 2,2',4,4',6,6'-Hexachlorobiphenyl | 0.108 | 6.54 | 6.67 | 6.51 |
| 156 | 2,3,3',4,4',5-Hexachlorobiphenyl | 0.119 | 7.44 | 6.67 | 7.07 |
| 158 | 2,3,3',4,4',6-Hexachlorobiphenyl | 0.115 | 6.78 | 6.67 | 6.87 |
| 163 | 2,3,3',4',5,6-Hexachlorobiphenyl | 0.115 | 6.78 | 6.67 | 6.87 |
| 164 | 2,3,3',4',5',6-Hexachlorobiphenyl | 0.110 | 6.63 | 6.67 | 6.61 |
| 165 | 2,3,3',5,5',6-Hexachlorobiphenyl | 0.115 | 7 | 6.67 | 6.87 |
| 167 | 2,3',4,4',5,5'-Hexachlorobiphenyl | 0.112 | 7.29 | 6.67 | 6.71 |
| 169 | 3,3',4,4',5,5'-Hexachlorobiphenyl | 0.112 | 7.55 | 6.67 | 6.71 |
| 170 | 2,2',3,3',4,4',5-Heptachlorobiphenyl | 0.116 | 7.08 | 7.08 | 7.09 |
| 172 | 2,2',3,3',4,5,5'-Heptachlorobiphenyl | 0.116 | 7.21 | 7.08 | 7.09 |
| 174 | 2,2',3,3',4,5,6'-Heptachlorobiphenyl | 0.113 | 6.85 | 7.08 | 6.94 |
| 175 | 2,2',3,3',4,5',6-Heptachlorobiphenyl | 0.115 | 6.92 | 7.08 | 7.04 |
| 176 | 2,2',3,3',4,6,6'-Heptachlorobiphenyl | 0.112 | 6.55 | 7.08 | 6.89 |
| 177 | 2,2',3,3',4,5',6'-Heptachlorobiphenyl | 0.114 | 6.73 | 7.08 | 6.99 |
| 178 | 2,2',3,3',5,5',6-Heptachlorobiphenyl | 0.114 | 6.85 | 7.08 | 6.99 |
| 179 | 2,2',3,3',5,6,6'-Heptachlorobiphenyl | 0.111 | 6.41 | 7.08 | 6.83 |
| 180 | 2,2',3,4,4',5,5'-Heptachlorobiphenyl | 0.115 | 7.21 | 7.08 | 7.04 |
| 181 | 2,2',3,4,4',5,6-Heptachlorobiphenyl | 0.119 | 7.13 | 7.08 | 7.24 |
| 182 | 2,2',3,4,4',5,6'-Heptachlorobiphenyl | 0.113 | 6.92 | 7.08 | 6.94 |
| 183 | 2,2',3,4,4',5',6-Heptachlorobiphenyl | 0.115 | 7.04 | 7.08 | 7.04 |
| 185 | 2,2',3,4,5,5',6-Heptachlorobiphenyl | 0.118 | 6.99 | 7.08 | 7.19 |
| 188 | 2,2',3,4',5,6,6'-Heptachlorobiphenyl | 0.111 | 6.78 | 7.08 | 6.83 |
| 189 | 2,3,3',4,4',5,5'-Heptachlorobiphenyl | 0.119 | 7.72 | 7.08 | 7.24 |
| 190 | 2,3,3',4,4',5,6-Heptachlorobiphenyl | 0.121 | 7.08 | 7.08 | 7.34 |
| 191 | 2,3,3',4,4',5',6-Heptachlorobiphenyl | 0.119 | 7.21 | 7.08 | 7.24 |
| 192 | 2,3,3',4,5,5',6-Heptachlorobiphenyl | 0.122 | 7.21 | 7.08 | 7.40 |
| 193 | 2,3,3',4',5,5',6-Heptachlorobiphenyl | 0.118 | 7.21 | 7.08 | 7.19 |
| 194 | 2,2',3,3',4,4',5,5'-Octachlorobiphenyl | 0.119 | 7.62 | 7.49 | 7.42 |
| 195 | 2,2',3,3',4,4',5,6-Octachlorobiphenyl | 0.121 | 7.35 | 7.49 | 7.52 |
| 196 | 2,2',3,3',4,4',5,6'-Octachlorobiphenyl | 0.118 | 7.43 | 7.49 | 7.36 |
| 203 | 2,2',3,4,4',5,5',6-Octachlorobiphenyl | 0.121 | 7.49 | 7.49 | 7.52 |
| 204 | 2,2',3,4,4',5,6,6'-Octachlorobiphenyl | 0.119 | 7.48 | 7.49 | 7.42 |
| 205 | 2,3,3',4,4',5,5',6-Octachlorobiphenyl | 0.125 | 7.62 | 7.49 | 7.72 |
| 206 | 2,2',3,3',4,4',5,5',6-Nonachlorobiphenyl | 0.124 | 7.94 | 7.89 | 7.84 |
| 207 | 2,2',3,3',4,4',5,6,6'-Nonachlorobiphenyl | 0.122 | 7.88 | 7.89 | 7.74 |

*Experimental data taken from References [19,20].
[a]Calculated log P values using one parameter ($N_{NH}$) regression.
[b]Calculated log P values using two parameter ($\omega$, $N_{NH}$) regression

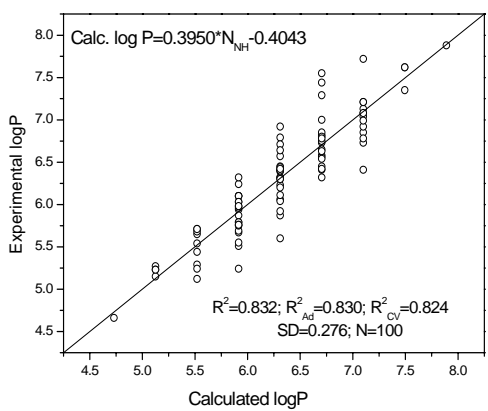 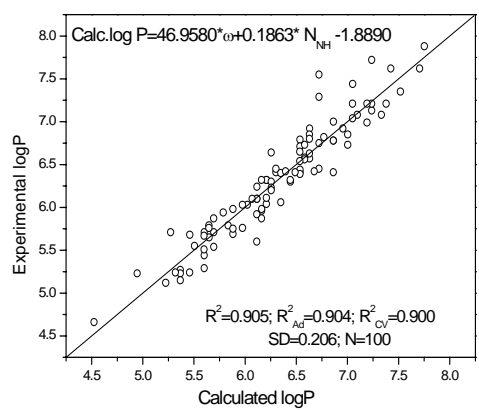

**(a)** **(b)**

**Figure 6:** Calculated vs experimental values of logP for the training set of 100 PCBs using (a) one parameter ($N_{NH}$) and (b) two parameter ($N_{NH}$, $\omega$) regressions

**Regression Equations:**

| Parameters | Regression Equations | $R^2$ | $R^2_{(CV)}$ | $R^2_{(Ad)}$ |
|---|---|---|---|---|
| One | Calculated log P=0.3950*$N_{NH}$-0.4043 | 0.832 | 0.824 | 0.830 |
| Two | Calculated log P=46.9580*$\omega$+0.1863* $N_{NH}$ -1.8890 | 0.905 | 0.900 | 0.904 |

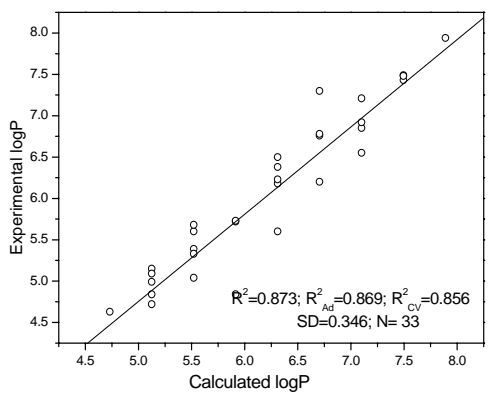 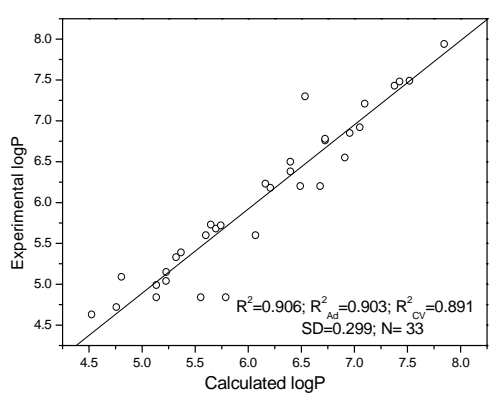

**(a)** **(b)**

**Figure 7:** Calculated vs experimental values of $\log P$ for the test set of 33 PCBs using (a) one parameter ($N_{NH}$) and (b) two parameter ($N_{NH}$, $\omega$) regressions.

**Regression Coefficients:**

| Parameters | $R^2$ | $R^2_{(CV)}$ | $R^2_{(Ad)}$ |
|---|---|---|---|
| **One** | 0.873 | 0.856 | 0.869 |
| **Two** | 0.906 | 0.891 | 0.903 |

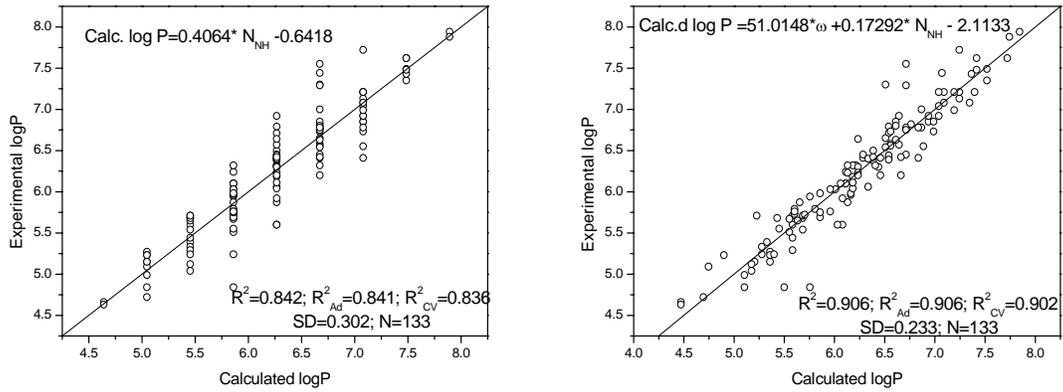

**Figure 8:** Calculated vs experimental values of $\log P$ for the complete set of 133 PCBs using (a) one parameter ($N_{NH}$) and (b) two parameter ($N_{NH}$, $\omega$) regressions.

**Regression Equations:**

| Parameters | Regression Equations | $R^2$ | $R^2_{(CV)}$ | $R^2_{(Ad)}$ |
|---|---|---|---|---|
| One | Calculated log P=0.4064* $N_{NH}$ -0.6418 | 0.842 | 0.836 | 0.841 |
| Two | Calculated log P =51.0148*$\omega$ +0.17292* $N_{NH}$ -2.1133 | 0.906 | 0.902 | 0.906 |

**Table 9:** QSPR models for log P calculation of chloroanisole

| Compounds | $\omega$ (au) | log P Value | | |
|---|---|---|---|---|
| | | Exptl* | Calcd[a] $N_{NH}$ | Calcd[b] $\omega$, $N_{NH}$ |
| 2-CAS | 0.0630 | 2.72 | 2.91 | 2.92 |
| 3-CAS | 0.0654 | 3.09 | 2.91 | 2.91 |
| 4-CAS | 0.0646 | 3.00 | 2.91 | 2.91 |
| 2,3-C2AS | 0.0753 | 3.30 | 3.45 | 3.47 |
| 2,4-C2AS | 0.0773 | 3.46 | 3.45 | 3.46 |
| 2,5-C2AS | 0.0770 | 3.27 | 3.45 | 3.46 |
| 2,6-C2AS | 0.0839 | 3.19 | 3.45 | 3.43 |
| 3,4-C2AS | 0.0768 | 3.55 | 3.45 | 3.46 |
| 3,5-C2AS | 0.0799 | 3.88 | 3.45 | 3.45 |
| 2,3,4-C3AS | 0.0871 | 3.92 | 4.00 | 4.02 |
| 2,3,5-C3AS | 0.0892 | 4.05 | 4.00 | 4.01 |
| 2,3,6-C3AS | 0.0943 | 3.76 | 4.00 | 3.99 |
| 2,4,5-C3AS | 0.0893 | 3.95 | 4.00 | 4.01 |
| 2,4,6-C3AS | 0.0977 | 4.05 | 4.00 | 3.97 |
| 3,4,5-C3AS | 0.0891 | 4.23 | 4.00 | 4.01 |
| 2,3,4,5-C4AS | 0.0991 | 4.57 | 4.55 | 4.57 |
| 2,3,4,6-C4AS | 0.1067 | 4.53 | 4.55 | 4.54 |
| 2,3,5,6-C4AS | 0.1057 | 4.52 | 4.55 | 4.54 |
| PCAS | 0.1162 | 5.16 | 5.10 | 5.10 |

*Experimental data taken from References [19,20].
[a]Calculated log P values using one parameter ($N_{NH}$) regression.
[b]Calculated log P values using two parameter ($\omega$, $N_{NH}$) regressio

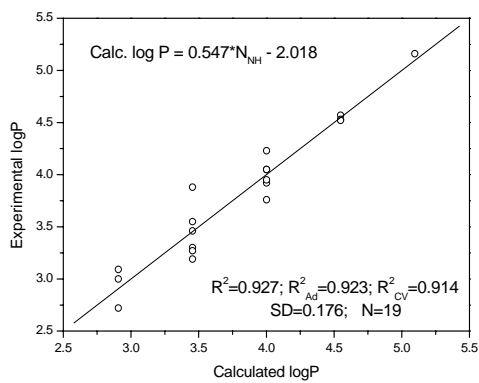 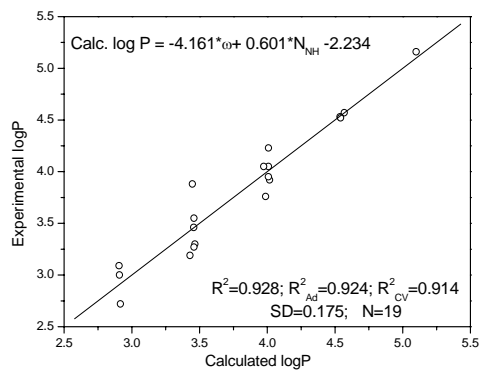

(a)            (b)

**Figure 9:** Calculated vs experimental values of log P of Chloroanisole using (a) one parameter ($N_{NH}$) and (b) two parameter ($N_{NH}$, $\omega$) regressions

**Regression Equations:**

| Parameters | Regression Equations | $R^2$ | $R^2_{(CV)}$ | $R^2_{(Ad)}$ |
|---|---|---|---|---|
| One | Calculated log P = 0.547*$N_{NH}$ - 2.018 | 0.927 | 0.914 | 0.923 |
| Two | Calculated log P = - 4.161*$\omega$ + 0.601* $N_{NH}$ -2.234 | 0.928 | 0.914 | 0.924 |

**Table 10:** p$K_a$ prediction using group philicity and no. of carbon atoms in a molecule.

| Molecules | $\omega_g^+$ | p$K_a$ Value | | |
| --- | --- | --- | --- | --- |
| | | Exptl* | Calcd[a] (N$_C$) | Calcd[b] ($\omega_g^+$, N$_C$) |
| *Acids* | | | | |
| pivalic acid | 0.612 | 5.05 | 4.68 | 5.01 |
| propionic acid | 0.661 | 4.87 | 3.45 | 4.18 |
| isobutyric acid | 0.618 | 4.86 | 4.06 | 4.67 |
| valeric acid | 0.653 | 4.86 | 4.68 | 4.84 |
| Butyric acid | 0.65 | 4.82 | 4.06 | 4.54 |
| isovaleric acid | 0.664 | 4.78 | 4.68 | 4.80 |
| Acetic acid | 0.694 | 4.76 | 2.83 | 3.72 |
| 4-bromobutyric acid | 0.713 | 4.58 | 4.06 | 4.28 |
| 4-chlorobutyric acid | 0.709 | 4.52 | 4.06 | 4.29 |
| glutaric acid | 0.718 | 4.35 | 4.68 | 4.57 |
| vinylacetic acid | 0.537 | 4.34 | 4.06 | 5.00 |
| succinic acid | 0.828 | 4.2 | 4.06 | 3.80 |
| 3-bromopropionic acid | 0.713 | 3.99 | 3.45 | 3.96 |
| Lactic acid | 0.713 | 3.86 | 3.45 | 3.96 |
| Glycolic acid | 0.699 | 3.83 | 2.83 | 3.70 |
| 2-hydroxybutyric acid | 0.74 | 3.68 | 4.06 | 4.17 |
| mercaptoacetic acid | 0.532 | 3.67 | 2.83 | 4.39 |
| Formic acid | 0.812 | 3.55 | 2.22 | 2.92 |
| 2-bromopropionic acid | 1.13 | 2.97 | 3.45 | 2.24 |
| bromoacetic acid | 0.77 | 2.9 | 2.83 | 3.41 |
| 2-chloropropionic acid | 1.053 | 2.88 | 3.45 | 2.56 |
| 2-chlorobutyric acid | 1.024 | 2.84 | 4.06 | 3.00 |
| malonic acid | 0.851 | 2.83 | 3.45 | 3.39 |
| Chloroacetic acid | 0.813 | 2.82 | 2.83 | 3.23 |
| fluoroacetic acid | 0.871 | 2.59 | 2.83 | 2.99 |
| 2-bromobutyric acid | 1.121 | 2.55 | 4.06 | 2.60 |
| cyanoacetic acid | 1.036 | 2.45 | 2.83 | 2.31 |
| dichloroacetic acid | 1.298 | 1.26 | 2.83 | 1.24 |
| difluroacetic acid | 1.141 | 1.24 | 2.83 | 1.88 |
| trichloroacetic acid | 1.389 | 0.63 | 2.83 | 0.86 |
| *Substituted Phenols* | | | | |
| o-methylphenol | 0.11 | 10.3 | 10.30 | 10.30 |
| o-chlorophenol | 0.15 | 8.6 | 8.43 | 9.21 |
| o-nitrophenol | 0.53 | 7.2 | 8.43 | 7.40 |
| m-nitrophenol | 0.49 | 8.4 | 8.43 | 7.59 |
| p-methoxyphenol | 0.11 | 10.2 | 10.30 | 10.30 |
| p-methylphenol | 0.12 | 10.3 | 10.30 | 10.20 |
| p-chlorophenol | 0.15 | 9.4 | 8.43 | 9.21 |
| p-nitrophenol | 0.44 | 7.2 | 8.43 | 7.83 |
| p-hydroxyphenol | 0.12 | 9.8 | 8.43 | 9.35 |
| *Alcohols* | | | | |
| 2-butanol | 0.194 | 17.6 | 17.02 | 17.68 |
| 1,2-ethanediol | 0.346 | 13.6 | 14.33 | 13.11 |
| 1,2-propanediol | 0.297 | 14.9 | 15.67 | 14.50 |
| 1,3- propanediol | 0.282 | 15.1 | 15.67 | 15.00 |
| 1,4-butanediol | 0.257 | 15.1 | 17.02 | 15.59 |
| Ethanol | 0.263 | 15.9 | 14.33 | 15.86 |
| propanol | 0.239 | 16.2 | 15.67 | 16.42 |
| tert-butyl alcohol | 0.181 | 19.2 | 17.02 | 18.11 |
| 2-methoxyethanol | 0.248 | 14.8 | 15.67 | 16.12 |

*Experimental data taken from References [24-26].

[a] Calculated p$K_a$ values using one parameter ($N_C$) regression.

[b] Calculated p$K_a$ values using two parameter ($\omega_g^+$, $N_C$) regression

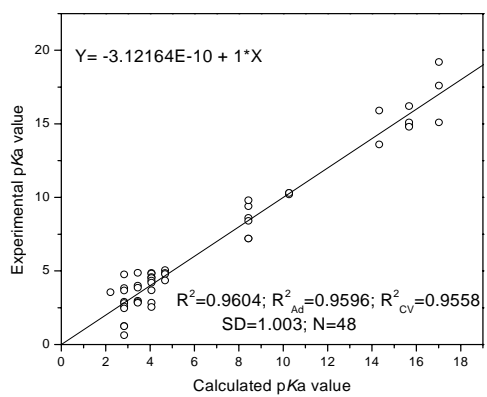 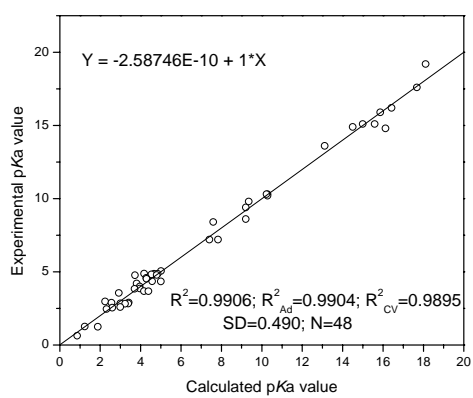

**(a)** **(b)**

**Figure 10:** Relationship between the calculated p$K_a$ values with experimental p$K_a$ values using a) one parameter ($N_C$) and b) two parameter ($\omega_g^+$, $N_C$) regressions

| Parameters | $R^2$ | $R^2$(cv) | $R^2$(ad) |
|---|---|---|---|
| **One** | 0.9604 | 0.9558 | 0.9596 |
| **Two** | 0.9906 | 0.9895 | 0.9904 |